\documentclass[amsmath,amssymb, twocolumn, superscriptaddress]{revtex4-2}
\usepackage{graphicx}
\usepackage{subcaption} 
\usepackage{caption}
\usepackage{dcolumn}
\usepackage{bm}
\usepackage{xcolor}
\usepackage{hyperref}
\hypersetup{
	colorlinks=true,
	linkcolor=blue, 
	citecolor=blue, 
	urlcolor=blue   
}
\usepackage[T1]{fontenc}
\usepackage[utf8]{inputenc}
\usepackage{miller}
\usepackage{comment}

\usepackage{threeparttable}
\usepackage{booktabs}
\usepackage{multirow}

\begin{document}

\title{\textbf{Benchmarking and advancing machine-learning interatomic potentials for 3C-SiC across equilibrium and extreme conditions} 
}%

\author{Ali Hamedani}
\email{Corresponding author; ali.hamedani@aalto.fi}
\affiliation{Department of Applied Physics, Aalto University, 00076 Aalto Espoo, Finland}%
\author{Zijian Meng}
\author{Laurent Karim B\'eland}
\affiliation{Department of Mechanical and Materials Engineering, Queen's University, Kingston, Ontario, K7L 3N6, Canada}%
\author{Andrea E. Sand}%
\affiliation{Department of Applied Physics, Aalto University, 00076 Aalto Espoo, Finland}%

\begin{abstract}
Several machine-learning interatomic potentials (MLIPs) have been developed for cubic silicon carbide (3C-SiC). Here, we benchmark seven existing MLIPs against a common set of structural, thermal, defect, radiation-damage, disordered-phase, and computational-performance tests. We also introduce two complementary MLIPs trained on DFT-labeled crystalline, liquid, amorphous, and defective configurations. TGAP, a Gaussian approximation potential based on the \verb|soap_turbo| descriptor, provides consistently high accuracy across the sampled configurational space. MTP, a custom moment tensor potential, retains good accuracy across these regimes while offering substantially lower computational cost, GPU acceleration, and access to multi-million-atom simulations. As an application-level test, both potentials remain stable in displacement cascades spanning short-range repulsion, transient disorder, and recrystallization. Overall, MTP emerges as a particularly strong accuracy--cost compromise for large-scale and strongly non-equilibrium simulations of 3C-SiC.
\end{abstract}

\maketitle

\section*{\label{sec:intro}Introduction}%

Silicon carbide (SiC) is a technologically important covalent ceramic with applications in photonics, microelectronics and extreme environments. Its strong directional bonding gives rise to an unusual combination of properties, including thermal-shock resistance \cite{r1}, hardness \cite{r2}, high thermal conductivity \cite{r3}, stability at elevated temperatures \cite{r4}, chemical resistance \cite{r5}, wear resistance \cite{r6}, and irradiation tolerance \cite{r7}. Among the many SiC polytypes, cubic 3C-SiC is widely studied as both a model system and a technologically relevant material for investigating thermal transport, high-temperature behavior, amorphization, irradiation damage, and defect formation. The mechanisms governing these phenomena often involve structural transformations and defect processes at atomic length and time scales that are difficult to resolve experimentally, motivating the use of atomistic simulation.

First-principles methods have been used extensively to characterize the mechanical and thermal response of carbide ceramics under extreme conditions~\cite{zrc1,zrc2}, but the accessible system sizes and time scales fall far below those required for processes such as melting, phase decomposition, and displacement cascades. Molecular dynamics (MD) simulations with interatomic potentials extend atomistic modeling to these larger spatial and temporal scales, although their usefulness depends critically on the balance between computational cost, accuracy, and transferability. Machine-learning interatomic potentials (MLIPs) provide a flexible framework for navigating this tradeoff by fitting interatomic interactions to reference electronic-structure data with model complexities that can be chosen according to the accuracy and computational efficiency required for a given application. Many MLIP formalisms are also systematically improvable, allowing accuracy to be increased through greater model complexity and broader training data, albeit generally at increased computational cost. This accuracy--complexity tradeoff has been demonstrated explicitly for MTPs, for which increasing model complexity improves predictions across structural, thermodynamic, and defect properties, while subsequent basis pruning can recover substantial computational efficiency~\cite{luo2023mtp,meng2026accelerating}. MTPs have also been applied to chemically and structurally diverse systems spanning crystalline and disordered phases, neutral and ionic states, and multicomponent liquids~\cite{zongo2024unified,sun2024interatomic,zongo2025amorphous,zongo2026machine}. A number of MLIPs have been proposed for 3C-SiC, including three Deep Potential models, hereafter referred to as DP-TH~\cite{dp-th}, DP-GB~\cite{dp-gb}, and DP-ZBL~\cite{dp-zbl}; two FLARE models, denoted here as FLARE-23~\cite{flare-23} and FLARE-26~\cite{flare-26}; a neural-network potential, denoted NNP~\cite{sic-nnp}; an ultra-fast force field, denoted UF3~\cite{uf3}; and a tabulated Gaussian approximation potential, denoted tabGAP~\cite{tabGAP}.

These models have substantially expanded the range of atomistic simulations accessible for SiC, but they were developed with different training domains, target applications, and validation criteria. Some primarily target crystalline configurations and near-equilibrium properties, including DP-TH, DP-GB, and FLARE-23. Others, such as FLARE-26, UF3, and NNP, extend their training domains to disordered configurations but do not explicitly include the broad range of defect environments relevant to radiation-damage simulations. DP-ZBL and tabGAP provide broader configurational coverage, incorporating crystalline, disordered, and defective structures. Their respective model constructions, however, occupy different points on the accuracy--cost spectrum. tabGAP uses a compact tabulated representation that provides high computational efficiency, but our benchmark, as we later present, shows larger deviations for some elastic properties and defect energetics. DP-ZBL employs a more flexible deep-learning representation and provides broad configurational coverage, but at a substantially higher computational cost, which can limit the system sizes and time scales accessible in large-scale simulations. More generally, because the available MLIPs have been trained and validated against different subsets of configurations and properties, comparisons reported across their original studies do not provide a common measure of their relative performance. Recent systematic benchmarking of interatomic potentials has similarly shown that good performance for conventional equilibrium properties does not necessarily imply comparable accuracy for more demanding quantities such as defect migration and short-range compression~\cite{thoms2025benchmark}. A systematic benchmark across the same set of equilibrium, defective, disordered, and highly non-equilibrium regimes is therefore needed to identify the accuracy--cost tradeoffs of the available models and to guide their selection for different applications.

In this work, we develop two complementary MLIPs for 3C-SiC using diverse DFT-labeled datasets spanning crystalline, liquid, and amorphous configurations, together with 21 types of point defects and defect complexes. The inclusion of both near-equilibrium and highly distorted atomic environments is intended to provide broad configurational coverage across the regimes relevant to thermal behavior, phase transformations, defect processes, and radiation damage. 

The first potential is a Gaussian approximation potential (GAP) trained using the many-body \verb|soap_turbo| descriptor~\cite{caro2019optimizing}, which we refer to as TGAP. TGAP is designed to prioritize accuracy across a broad range of 3C-SiC environments. The second is a moment tensor potential, denoted MTP, that departs from the original MTP formalism~\cite{MTP} in its level definitions and basis-set construction, as described below. MTP is trained for the Kokkos-accelerated implementation that we recently introduced~\cite{mtp-kokkos}, providing GPU support and efficient access to large system sizes. The two models therefore occupy complementary positions on the accuracy--cost spectrum: TGAP emphasizes accuracy across the sampled configurational space, whereas MTP emphasizes computational efficiency while retaining good accuracy across the targeted regimes. 

Alongside the development of TGAP and MTP, we benchmark seven existing MLIPs and selected classical potentials for 3C-SiC using a common set of structural, mechanical, thermal, defect, disordered-phase, radiation-damage, and computational-performance tests. Applying the same benchmarks across models provides a consistent basis for comparing their strengths, limitations, and accuracy--cost tradeoffs. 

Finally, we use displacement cascades as an application-level test of the two developed potentials. A single cascade samples several qualitatively distinct regions of the potential-energy surface, including near-equilibrium crystalline environments, strongly repulsive short-range interactions, transiently disordered and melt-like configurations, and subsequent recrystallization into a defective lattice. Displacement cascades therefore provide a demanding test of model stability and of the ability to describe multiple configurational regimes within a single simulation, while also demonstrating the computational scales accessible with the two potentials.

\section*{\label{sec:ml_models}ML model construction}

The GAP framework was selected in part because of our existing computational ecosystem for radiation-damage simulations with GAP-based MLIPs. At its core is TurboGAP~\cite{caro2019turbogapwebsite}, a molecular dynamics code developed specifically for efficient simulations with GAP models. We recently extended TurboGAP with functionality required for radiation-damage simulations, including capabilities for treating the highly non-equilibrium conditions encountered during displacement cascades~\cite{Si_TurboGAP}. GAP also remains competitive with more recent MLIP formalisms in terms of data efficiency and achievable accuracy~\cite{bench1,bench2}. The \verb|soap_turbo| descriptor used here is an optimized variant of the smooth overlap of atomic positions (SOAP) representation~\cite{SOAP}, designed to reduce the computational cost of descriptor evaluation through optimized radial basis functions and related numerical improvements~\cite{caro2019optimizing,wang2022structure}. We refer to the resulting potential as TGAP.

The MTP developed here builds on the moment tensor potential formalism introduced by Shapeev~\cite{MTP} and implemented in the MLIP-2 and MLIP-3 packages~\cite{novikov2021mlip,podryabinkin2023mlip}, as well as on our previous development and application of MTPs across increasingly complex configurational domains~\cite{luo2023mtp,zongo2024unified,zongo2025amorphous}. In particular, our unified Si/O/SiO$_2$ MTP demonstrated that a single MTP can describe crystalline structures, point and extended defects, liquid and amorphous phases, interfaces, melting, and defect migration within a common framework~\cite{zongo2024unified}. The present model uses a modified definition of model complexity and basis-set construction. Alternative MTP level definitions have previously been explored~\cite{wang2025efficient}; here, we extend this approach through empirical optimization of the basis construction. The final model contains 9527 basis functions and uses modified level constants together with explicit limits on the number of contracted tensors and the maximum moment-tensor rank. This construction yields an inference cost comparable to that of a conventional level-18 MTP while achieving substantially lower fitting errors than a level-28 model. Further details and quantitative comparisons are provided in Supplementary Note~2. Training was performed using a custom adaptation of MLIP-3~\cite{podryabinkin2023mlip}, with modifications to the numerical linear-algebra routines to support the enlarged basis set. Post-training basis pruning was subsequently applied to further improve the accuracy--cost tradeoff~\cite{meng2026accelerating}.

The initial training dataset, used for TGAP and as the basis for MTP training, contains both crystalline and disordered configurations (Supplementary Note~1). The crystalline configurations sample several classes of atomic environments. (i) Elastically deformed structures with different degrees of expansion and compression sample bulk response and the local curvature of the potential-energy surface. (ii) Configurations spanning a range of lattice constants and atomic-displacement amplitudes sample thermal and vibrational environments. (iii) Defective structures provide coverage of point defects and defect complexes. In total, 21 defect types are included: two vacancies, one divacancy, two trivacancies, two antisites, four tetrahedral interstitials, two hexagonal interstitials, four \hkl[100] split interstitials, and four \hkl[110] split interstitials.

The disordered portion of the dataset was initially generated from \textit{ab initio} molecular dynamics (AIMD) melt--quench simulations. The emphasis on explicit coverage of crystalline, thermally distorted, defective, and disordered environments follows the approach of our earlier work on Si~\cite{hamedani2021primary, hamedani2020insights} and of general-purpose machine-learning potentials more broadly~\cite{gap_C,gap_P,volker-silica,GeSbTe-volker}. It also builds on our previous experience developing MTPs for Si and the Si/O/SiO$_2$ system~\cite{zongo2022database,zongo2024unified}. Configurations were sampled during high-temperature randomization at 6000~K, equilibration at density-dependent temperatures, intermediate stages of quenching to 300~K, and the resulting amorphous state. Six densities ranging from 2.3 to 3.2~g/cm$^3$ were considered, with the upper limit corresponding approximately to the density of crystalline zinc-blende SiC. Additional disordered configurations for TGAP were collected iteratively from TGAP-driven melt--quench trajectories. After a stable TGAP dataset had been established, the MTP dataset was further augmented using active learning to improve stability in disordered environments. An initial MTP trained on the TGAP dataset was used to identify configurations entering the extrapolative regime, based on the uncertainty measure intrinsic to the MTP formalism, and selected configurations were subsequently added to the training set. Related small-cell and active-learning strategies developed in our previous work have shown that targeted selection of informative configurations can substantially reduce the DFT cost of constructing robust MLIPs~\cite{meziere2023accelerating,meng2025smallcell}. Details of the present active-learning procedure and the number of configurations in each category are provided in Supplementary Note~1 and Supplementary Table~1, respectively.

The hyperparameters used for TGAP and MTP training are reported in Supplementary Note~2, numerical validation of the fitted models is presented in Supplementary Note~3, and the computational details of DFT labeling and validation simulations are provided in Supplementary Note~4.

\section*{\label{sec:res}Results}%
\subsection{Bulk and thermal properties}

\begin{figure}[htbp!]
	\centering
	\includegraphics[width=1.0\columnwidth]{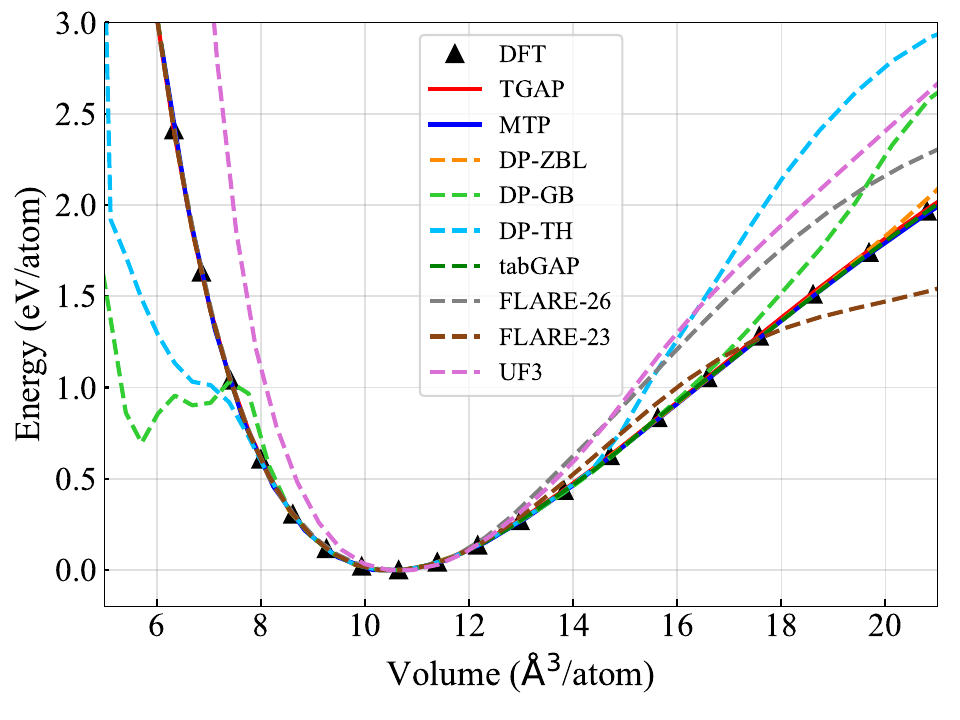}
	\caption{Energy--volume curves of 3C-SiC calculated with the MLIPs and PBE-DFT.}
	\label{fig:e-v}
\end{figure}

Figure~\ref{fig:e-v} compares the energy--volume relationship of 3C-SiC predicted by TGAP and MTP with DFT and the other MLIPs. TGAP, MTP, tabGAP, and DP-ZBL closely reproduce the DFT curve across the full volume range considered, whereas the remaining models show larger deviations under compression, expansion, or both.

The energy--volume relationship provides an initial benchmark of how accurately each potential describes the response of the crystal to volumetric deformation away from equilibrium. Accurate reproduction of the DFT curve is particularly important for simulations involving substantial compression or expansion and provides a baseline test of the potential-energy surface beyond the equilibrium structure. Energy--volume curves obtained with the classical potentials are presented in Supplementary Note~5.

Table~\ref{tab:bulk} summarizes the equilibrium structural and elastic properties of 3C-SiC predicted by TGAP and MTP, together with the other MLIPs, our DFT calculations, literature DFT values, and available experimental measurements.

\begin{table*}[ht]
	\centering
	\caption{Equilibrium lattice constant $a$ (in \AA), elastic constants $C_{ij}$ (in GPa), Voigt--Reuss--Hill averages of the bulk modulus $B$ and shear modulus $G$ (in GPa), Young's modulus $E$ (in GPa), and Poisson's ratio $\nu$ for 3C-SiC. Results from TGAP, MTP, and the other MLIPs are compared with our PBE-DFT calculations, literature DFT values, and experimental measurements. Our DFT elastic constants were calculated using density-functional perturbation theory (DFPT) as implemented in VASP.}
	\label{tab:bulk}
	\begin{ruledtabular}
		\renewcommand{\arraystretch}{1.3} 
		\begin{tabular}{lcccccccccccc}
			& TGAP & MTP & DP-ZBL & tabGAP & DP-GB & DP-TH & FLARE-23 & FLARE-26 & UF3 & DFT & $\mathrm{DFT}_{\mathrm{lit.}}$\textsuperscript{a} & Exp. \\
			\hline
			$a$ (\AA) & 4.3769 & 4.3776 & 4.3799 & 4.3811 & 4.3789 & 4.3789 & 4.3792 & 4.3917 & 4.4049 & 4.3788 & 4.3805 & 4.3596\textsuperscript{a} \\
			$C_{11}$ & 374.8 & 385.1 & 381.8 & 309.0 & 381.6 & 388.7 & 373.0 & 434.1 & 433.6 & 384.5 & 384.3 & 390\textsuperscript{b}, 395\textsuperscript{c}, 371\textsuperscript{d} \\
			$C_{12}$ & 125.0 & 128.3 & 147.1 & 173.7 & 127.5 & 120.3 & 128.1 & 106.4 & 119.4 & 127.2 & 127.9 & 142\textsuperscript{b}, 123\textsuperscript{c}, 146\textsuperscript{d} \\
			$C_{44}$ & 231.8 & 237.0 & 228.8 & 209.9 & 238.5 & 240.1 & 230.3 & 238.8 & 193.4 & 241.1 & 239.9 & 150\textsuperscript{e}, 256\textsuperscript{b}, 236\textsuperscript{c}, 111\textsuperscript{d} \\
			$B$ & 208.3 & 213.9 & 225.3 & 218.8 & 212.2 & 209.8 & 209.7 & 215.6 & 224.1 & 213.0 & 213.6 & 225\textsuperscript{b}, 270\textsuperscript{e} \\
			$G$ & 180.9 & 185.3 & 175.0 & 133.5 & 185.2 & 190.1 & 178.7 & 205.3 & 178.0 & 187.4 & 186.7 & 192\textsuperscript{f} \\
			$E$ & 420.8 & 431.4 & 417.1 & 332.8 & 430.4 & 438.0 & 417.6 & 467.6 & 422.1 & 434.6 & 433.7 & 448\textsuperscript{f} \\
			$\nu$ & 0.163 & 0.164 & 0.192 & 0.246 & 0.162 & 0.152 & 0.168 & 0.139 & 0.186 & 0.160 & 0.162 & 0.267\textsuperscript{b}, 0.168\textsuperscript{f}
		\end{tabular}
		\footnotetext{Ref.~\cite{prb_polytypes}}
		\footnotetext{Ref.~\cite{exp1}}
		\footnotetext{Ref.~\cite{exp2}}
		\footnotetext{Ref.~\cite{exp3}}
		\footnotetext{Ref.~\cite{exp4}}
		\footnotetext{Ref.~\cite{exp5}}
	\end{ruledtabular}
\end{table*}

Our DFT elastic constants were calculated within density-functional perturbation theory (DFPT)~\cite{dfpt1,dfpt2}, using VASP~\cite{vasp1} and the PBE exchange--correlation functional~\cite{pbe}. Both TGAP and MTP reproduce the equilibrium lattice constant closely. MTP also gives excellent agreement with the DFT elastic constants, differing by less than 5~GPa for each of $C_{11}$, $C_{12}$, and $C_{44}$. TGAP reproduces $C_{12}$ closely and underestimates $C_{11}$ and $C_{44}$ by approximately 10~GPa. The resulting Voigt--Reuss--Hill bulk, shear, and Young's moduli remain close to the DFT values for both models.

Among the other MLIPs, DP-GB provides particularly uniform agreement with DFT, reproducing all three elastic constants to within 3~GPa. DP-TH similarly reproduces $C_{11}$ and $C_{44}$ closely but underestimates $C_{12}$. Other models show larger deviations in particular components: tabGAP substantially underestimates $C_{11}$ and consequently the shear and Young's moduli, whereas UF3 overestimates $C_{11}$ and underestimates $C_{44}$. Overall, MTP, TGAP, and several existing MLIPs accurately reproduce the equilibrium structure of 3C-SiC, but the comparison reveals appreciable differences in how uniformly they describe its elastic response. The experimental values span a comparatively broad range for several quantities, so the DFT calculations provide the more consistent reference for comparing the models.

We next examined the temperature dependence of Young's modulus, for which experimental measurements have been summarized in Ref.~\cite{handbook}. Molecular dynamics simulations were performed from 100 to 2000~K using 1000-atom supercells. The finite-temperature elastic constants were calculated from the stress-fluctuation and Born contributions~\cite{Born_matrix}, as implemented in LAMMPS~\cite{LAMMPS}, and were used to obtain the corresponding Young's modulus. No quantum corrections~\cite{QEffect} were applied to the MLIP results.

Ref.~\cite{handbook} represents the experimentally measured temperature dependence of Young's modulus using the empirical relation
\begin{equation}
	\label{eq:ET}
	E(T)=E_0-BT\exp\left(-\frac{T_0}{T}\right),
\end{equation}
where $E(T)$ is Young's modulus at temperature $T$, and $E_0$, $B$, and $T_0$ are fitting parameters. The experimental fit reported in Ref.~\cite{handbook} uses $E_0=460$~GPa, $B=0.040$~GPa/K, and $T_0=962$~K. For comparison, we fitted Eq.~\ref{eq:ET} independently to the Young's moduli obtained with each MLIP over the range 100--2000~K. The resulting parameters are listed in Table~\ref{tab:elastic-fit-params}.

\begin{table}[!htbp]
	\centering
	\begin{threeparttable}
		\caption{Parameters obtained by fitting Eq.~\ref{eq:ET} to the temperature-dependent Young's modulus predicted by each MLIP. The experimental parameters are those reported in Ref.~\cite{handbook}.}
		\label{tab:elastic-fit-params}
		\renewcommand{\arraystretch}{1.3}
		\begin{tabular}{lccc}
			\toprule
			Potential & $E_0$ (GPa) & $B$ (GPa/K) & $T_0$ (K) \\
			\midrule
			TGAP     & 417 & 0.041 & 737 \\
			MTP      & 430 & 0.030 & 272 \\
			tabGAP   & 635 & 0.102 & 0 \\
			FLARE-26 & 464 & 0.044 & 277 \\
			UF3      & 462 & 0.044 & 163 \\
			DP-ZBL   & 413 & 0.035 & 0 \\
			DP-GB    & 430 & 0.032 & 151 \\
			DP-TH    & 435 & 0.036 & 250 \\
			\midrule
			Exp.     & 460 & 0.040 & 962 \\
			\bottomrule
		\end{tabular}
	\end{threeparttable}
\end{table}

\begin{figure}[htbp!]
	\centering
	\includegraphics[width=1.0\columnwidth]{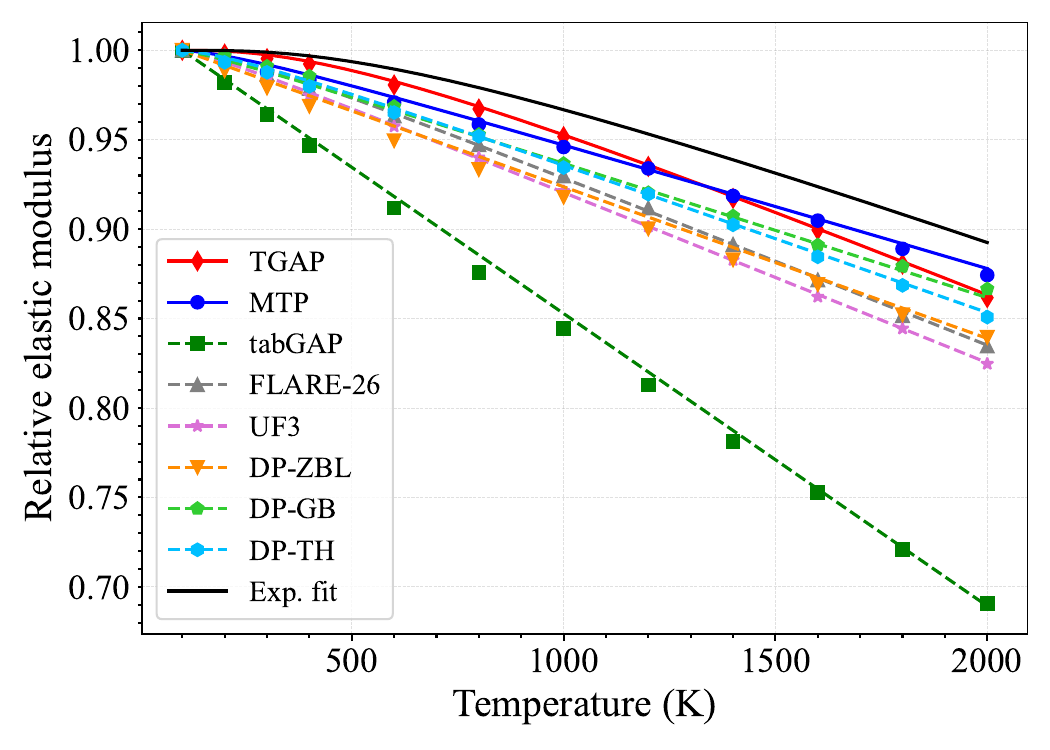}
	\caption{Relative Young's modulus, $E(T)/E(100~\mathrm{K})$, obtained from MD simulations with the MLIPs. The black solid line is the correspondingly normalized experimental relation $E(T)=460-0.04\,T\exp(-962/T)$ reported in Ref.~\cite{handbook}. The parameters obtained by fitting the same functional form to each MLIP are listed in Table~\ref{tab:elastic-fit-params}.}
	\label{fig:YT}
\end{figure}

Figure~\ref{fig:YT} compares the relative Young's modulus, $E(T)/E(T_{\mathrm{ref}})$, predicted by the MLIPs with the empirical experimental trend, using $T_{\mathrm{ref}}=100$~K for normalization.

All models predict a monotonic decrease in Young's modulus with increasing temperature, consistent with the experimental trend. TGAP and MTP provide the closest overall agreement with the normalized experimental curve across the investigated temperature range. TGAP follows the experimental trend more closely at lower and intermediate temperatures, whereas the TGAP and MTP curves cross near 1300~K and MTP shows the smaller deviation at higher temperatures. DP-GB and DP-TH show intermediate agreement, while FLARE-26, DP-ZBL, and UF3 generally predict stronger thermal softening. tabGAP exhibits the strongest reduction in relative Young's modulus at high temperature.

The fitted parameters in Table~\ref{tab:elastic-fit-params} provide an additional compact description of these trends. TGAP gives values of $B$ and $T_0$ closest to the experimental fit, whereas the remaining models yield substantially smaller $T_0$ values. Because the three fitting parameters are correlated and, for some models, the optimum occurs at $T_0=0$, the fitted parameters should be interpreted primarily as empirical descriptors of the simulated temperature dependence rather than as independent physical quantities.

\begin{figure}[htbp!]
	\centering
	\includegraphics[width=1.0\columnwidth]{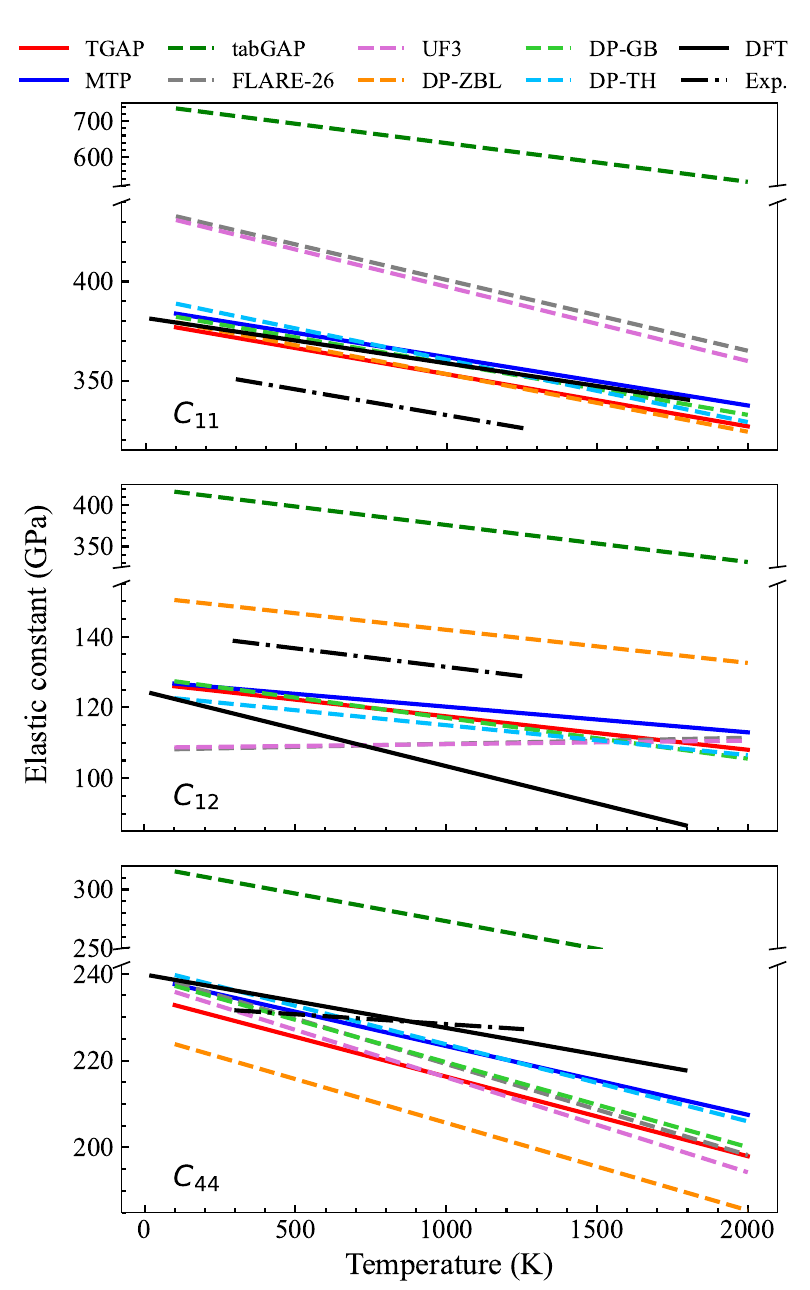}
	\caption{Temperature dependence of the elastic constants $C_{11}$, $C_{12}$, and $C_{44}$ of 3C-SiC obtained from MD simulations with the MLIPs, compared with DFT~\cite{elcnts_DFT} and experimental~\cite{elcnts_exp} reference data. Broken $y$ axes are used to preserve the resolution of the main group of predictions while displaying the full range of values. Slopes obtained from linear fits to each dataset are listed in Table~\ref{tab:Cij-slopes}.}
	\label{fig:elas-vs-temp}
\end{figure}

Figure~\ref{fig:elas-vs-temp} shows the temperature dependence of the elastic constants $C_{11}$, $C_{12}$, and $C_{44}$ predicted by TGAP, MTP, and the other MLIPs, together with DFT~\cite{elcnts_DFT} and experimental~\cite{elcnts_exp} reference data. Linear fits, $C_{ij}(T)=mT+b$, were used to quantify the temperature dependence of each elastic constant, with the fitted slopes reported in Table~\ref{tab:Cij-slopes}.

\begin{table}[!htbp]
	\centering
	\captionsetup{width=0.60\columnwidth, justification=justified, singlelinecheck=false, skip=2pt}
	\begin{threeparttable}
		\caption{Slopes obtained from linear fits to the temperature-dependent elastic constants, $C_{ij}(T)=mT+b$. The slopes $m$ are given in GPa/K.}
		\label{tab:Cij-slopes}
		\renewcommand{\arraystretch}{1.3}
		\begin{tabular}{lccc}
			\toprule
			Potential & $C_{11}$ & $C_{12}$ & $C_{44}$ \\
			\midrule
			TGAP     & $-0.0263$ & $-0.0094$ & $-0.0183$ \\
			MTP      & $-0.0244$ & $-0.0073$ & $-0.0158$ \\
			tabGAP   & $-0.1076$ & $-0.0449$ & $-0.0469$ \\
			FLARE-26 & $-0.0357$ & $ 0.0017$ & $-0.0210$ \\
			UF3      & $-0.0375$ & $ 0.0010$ & $-0.0219$ \\
			DP-ZBL   & $-0.0291$ & $-0.0094$ & $-0.0201$ \\
			DP-GB    & $-0.0260$ & $-0.0115$ & $-0.0195$ \\
			DP-TH    & $-0.0315$ & $-0.0085$ & $-0.0177$ \\
			\midrule
			DFT      & $-0.0229$ & $-0.0211$ & $-0.0123$ \\
			Exp.     & $-0.0259$ & $-0.0105$ & $-0.0045$ \\
			\bottomrule
		\end{tabular}
	\end{threeparttable}
\end{table}

All models predict a decrease in $C_{11}$ and $C_{44}$ with increasing temperature, consistent with the thermal softening of the crystal. For $C_{11}$, MTP most closely reproduces the DFT temperature dependence, with a fitted slope of $-0.0244$~GPa/K compared with $-0.0229$~GPa/K from DFT. TGAP and DP-GB also reproduce the temperature dependence closely, with DP-GB nearly matching the experimental slope of $-0.0259$~GPa/K. In terms of the absolute values of $C_{11}$, DP-GB remains closest to DFT over much of the temperature range, while TGAP, MTP, DP-TH, and DP-ZBL form a relatively narrow group around the DFT reference. FLARE-26 and UF3 predict systematically higher values, and tabGAP shows both substantially higher values and a much stronger temperature dependence. All of the MLIPs overestimate $C_{11}$ relative to the experimental reference.

The behavior of $C_{12}$ illustrates the differences between the DFT and experimental temperature trends. TGAP, MTP, DP-GB, and DP-TH lie close to the DFT values at low temperature but exhibit substantially weaker temperature dependence than the DFT reference. Their fitted slopes range from $-0.0073$ to $-0.0115$~GPa/K, close to the experimental slope of $-0.0105$~GPa/K but considerably smaller in magnitude than the DFT value of $-0.0211$~GPa/K. Consequently, these models agree well with DFT near the lower end of the temperature range but progressively deviate from it as the temperature increases. DP-ZBL predicts somewhat larger $C_{12}$ values and lies closer to the experimental reference. FLARE-26 and UF3 instead predict a nearly temperature-independent $C_{12}$ and therefore reproduce the softening trend of neither DFT nor experiment, while tabGAP shows substantially stronger softening than either reference.

For $C_{44}$, MTP gives the temperature dependence closest to DFT, followed by DP-TH and TGAP. The absolute predictions from these models also remain relatively close to the DFT reference, whereas DP-ZBL gives lower values and tabGAP systematically higher values. In contrast to $C_{11}$ and $C_{12}$, however, none of the MLIPs reproduces the weak experimental temperature dependence of $C_{44}$. The experimental slope is only $-0.0045$~GPa/K, whereas the MLIP slopes range from $-0.0158$ to $-0.0219$~GPa/K for the main group of models, with tabGAP showing an even stronger decrease.

Overall, the finite-temperature elastic response is reproduced with varying accuracy depending on the elastic component and the chosen reference. TGAP, MTP, DP-GB, and DP-TH provide comparatively balanced descriptions across the three elastic constants, but no model reproduces both the DFT and experimental temperature dependence uniformly. In particular, the consistently excessive softening of $C_{44}$ relative to experiment remains a common limitation of the tested MLIPs.

\begin{figure*}[htbp!]
	\centering
	\includegraphics[width=0.95\textwidth]{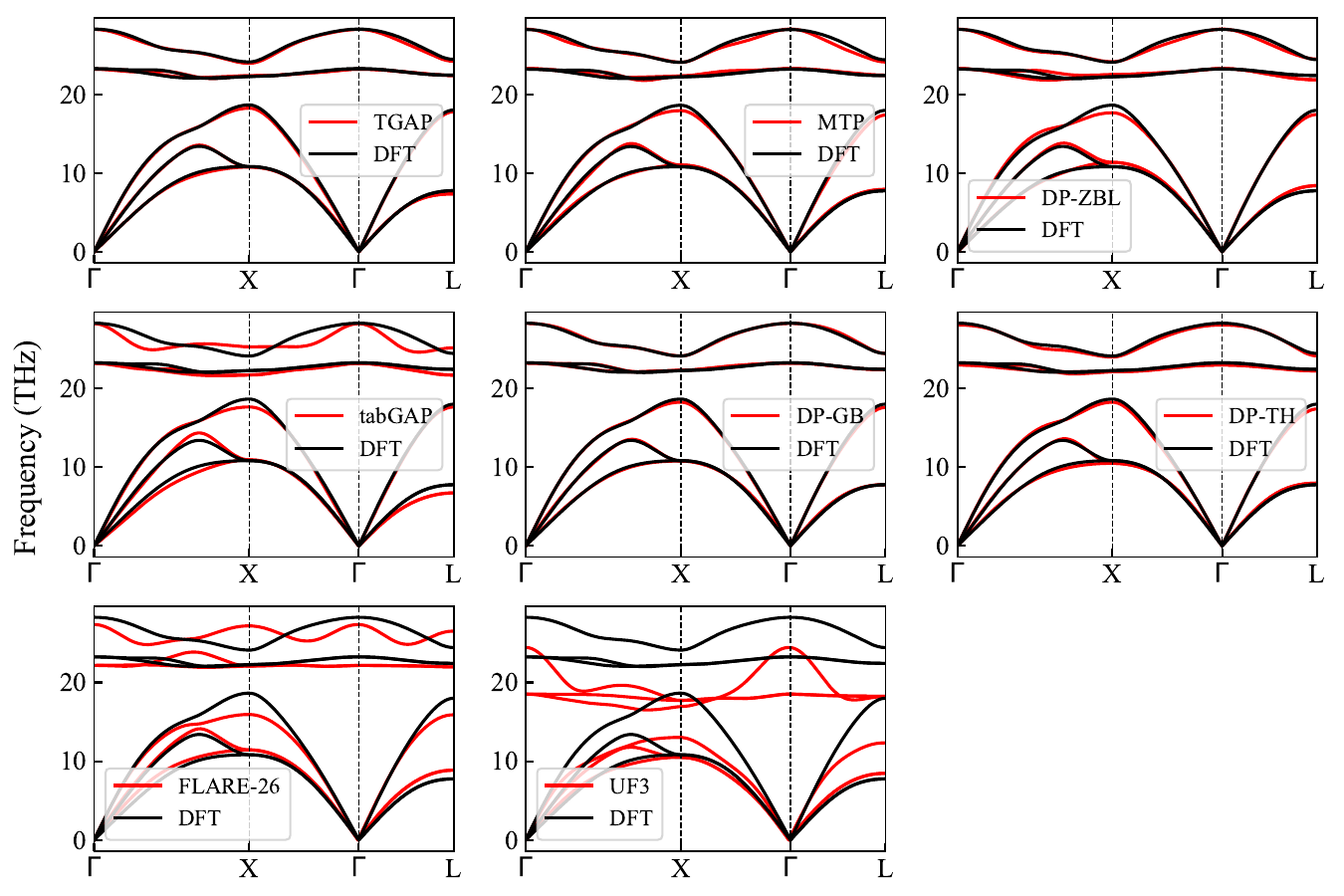}
	\caption{Phonon dispersion curves of 3C-SiC along the $\Gamma \to \mathrm{X} \to \Gamma \to \mathrm{L}$ path in reciprocal space, calculated using the MLIPs and DFT. Phonon calculations were performed with Phonopy using 128-atom supercells constructed at the equilibrium lattice constant of each model.}
	\label{fig:phonon}
\end{figure*}

Vibrational properties probe the local curvature of the potential-energy surface around equilibrium configurations and provide an important test of the forces governing thermodynamic and transport behavior. We calculated the phonon dispersion of 3C-SiC using Phonopy~\cite{phonopy} and compare the MLIP predictions with DFT in Figure~\ref{fig:phonon}.

TGAP, MTP, and the Deep Potential models closely reproduce the DFT phonon dispersion, including both the acoustic and optical branches. tabGAP also shows good overall agreement, although deviations are visible in the upper optical modes. FLARE-26 shows larger discrepancies, particularly in the optical branches and in portions of the acoustic spectrum. UF3 exhibits the largest deviations among the tested MLIPs, most notably in the optical branches, where the dispersion is substantially weaker than in DFT, with smaller deviations also present in the acoustic modes. The phonon dispersions obtained with the classical potentials are presented in Supplementary Note~6.

\begin{figure}[htbp!]
	\centering
	\includegraphics[width=0.99\columnwidth]{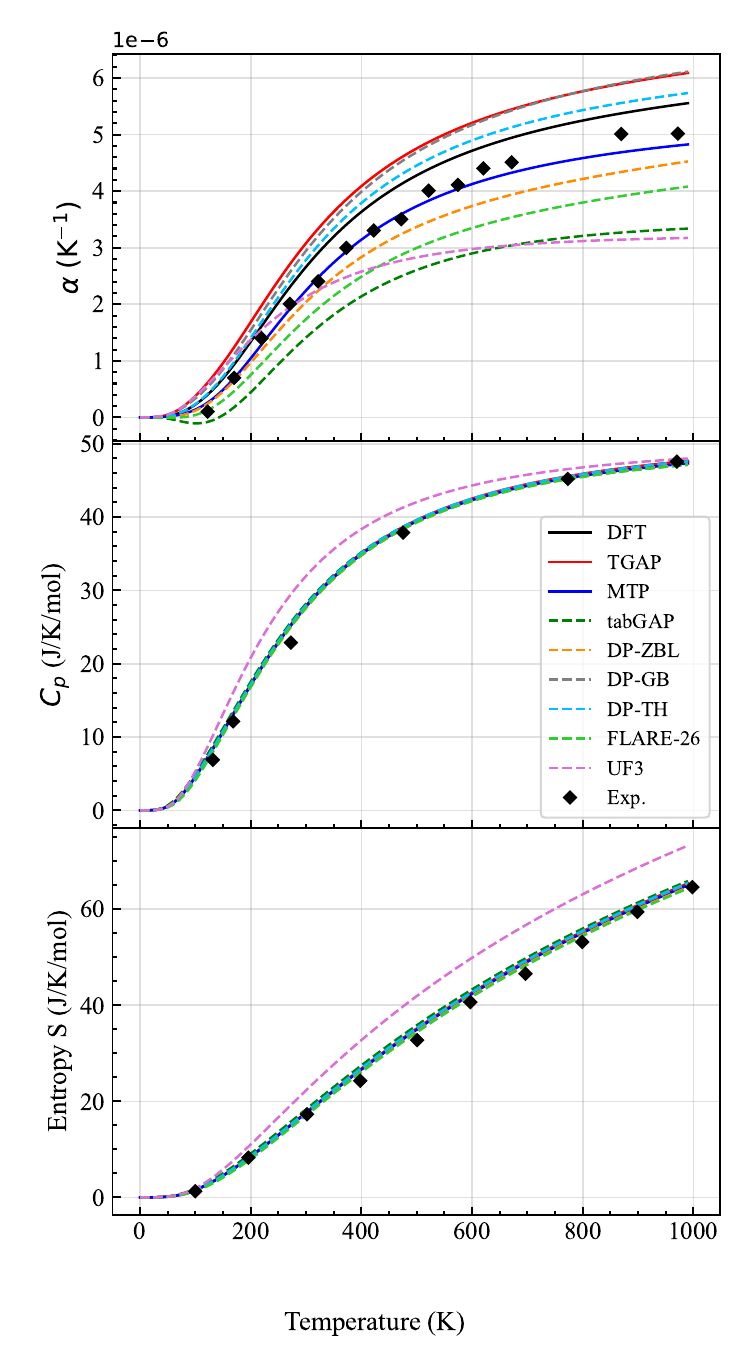}
	\caption{Linear thermal expansion coefficient, heat capacity, and entropy of 3C-SiC calculated within the quasi-harmonic approximation (QHA) using DFT and the MLIPs. The QHA calculations were performed with Phonopy using lattice distortions of up to 5\%. Experimental data are taken from Ref.~\cite{handbook}.}
	\label{fig:qha}
\end{figure}

We also calculated the linear thermal expansion coefficient, heat capacity, and entropy within the quasi-harmonic approximation (QHA) using Phonopy. Figure~\ref{fig:qha} compares the temperature dependence of these quantities predicted by the MLIPs with DFT and experimental data from Ref.~\cite{handbook}.

For the heat capacity and entropy, most MLIPs closely reproduce the DFT predictions and agree well with the experimental data over the investigated temperature range. The main exception is UF3, which overestimates $C_\mathrm{p}$ below approximately 800~K, with the largest deviation occurring between 200 and 400~K. At higher temperatures, its heat capacity converges toward the other predictions. The corresponding difference accumulates in the entropy, leading UF3 to progressively overestimate the DFT and experimental values with increasing temperature.

Larger differences among the models are observed for the linear thermal expansion coefficient. DP-TH follows the DFT prediction most closely over the investigated temperature range. MTP predicts somewhat smaller thermal expansion than DFT but lies closest to the experimental measurements. TGAP and DP-GB predict larger expansion than DFT at intermediate and high temperatures, whereas DP-ZBL, FLARE-26, UF3, and tabGAP predict smaller values to varying degrees. Thus, the model that most closely reproduces the DFT thermal expansion is not the one that agrees best with experiment, reflecting the difference between the DFT and experimental reference curves themselves.

Overall, TGAP and MTP reproduce the DFT heat capacity and entropy closely, together with most of the tested MLIPs. The thermal expansion coefficient provides a more discriminating benchmark, with MTP giving the closest agreement with the available experimental data.

\subsection{Repulsive response}

Radiation-damage simulations require interatomic potentials to describe not only near-equilibrium bonding, but also the strongly repulsive interactions encountered when energetic atoms approach one another at short distances. Previous cascade simulations have shown that energetic collisions can generate local pressures exceeding tens of GPa and interatomic separations below 2~\AA, and that the treatment of these short-range interactions can materially affect the predicted production of stable radiation damage~\cite{San15a.beland2016extreme,beland2017shortrange}. More recently, short-range reparameterization of Zr potentials was shown to substantially alter predicted threshold displacement energies while leaving many near-equilibrium properties comparatively unchanged~\cite{ghorbani2025shortdistance}. To provide this short-range behavior, TGAP and MTP were coupled to the Nordlund--Lehtola--Hobler (NLH) repulsive potential~\cite{NLH}, $V_\mathrm{NLH}$. The NLH screening function was fitted to all-electron electronic-structure calculations performed with numerical atomic orbitals using DMOL97. These calculations were benchmarked against second-order M\o{}ller--Plesset perturbation theory (MP2) calculations with flexible Gaussian-type orbital basis sets~\cite{NLH}. Relative to the commonly used ZBL potential~\cite{zbl}, the NLH potential provides improved agreement with these benchmark short-range interaction energies. The functional form of $V_\mathrm{NLH}$ and the parameters used here are given in Supplementary Note~7.

In the GAP framework, the repulsive contribution can be incorporated directly as a baseline potential~\cite{volker_gaussian}. During training, the baseline energy and forces are subtracted from the reference data and the residual contribution is fitted by the GAP model. At inference, the fitted residual is added back to the baseline. This construction allows the explicitly repulsive short-range interaction to be combined smoothly with the many-body MLIP description at larger separations.

For MTP, incorporation of the repulsive interaction required modification of the radial basis. The commonly used \texttt{RBChebyshev} basis employs a quadratic envelope that is $C^1$ continuous at the cutoff distance, but polynomial extrapolation below its specified minimum distance, $R_\mathrm{min}$, can lead to poorly controlled behavior in the short-range regime. We therefore introduced a custom radial basis with an envelope that is $C^2$ continuous at both $R_\mathrm{min}$ and the outer cutoff. The envelope is constructed from two smootherstep functions joined at a user-defined peak position selected from the pair-distance distribution of the training set. This provides a controlled transition between the MTP representation and the short-range repulsive interaction. Related hybridization strategies have been used in Deep Potential models~\cite{wang2019deep,dp-zbl} and atomic cluster expansion models~\cite{bochkarev2022efficient}.

The training dataset includes isolated Si--Si, Si--C, and C--C dimers spanning interatomic separations of 1.3--7.0, 1.1--7.0, and 0.9--7.0~\AA, respectively. At separations below the lower bound of each range, the interaction is governed by $V_\mathrm{NLH}$. The lower bounds were chosen based on the magnitude of the corresponding forces and the range over which the reference DFT calculations remain reliable.

\begin{figure*}[htbp!]
	\centering
	\includegraphics[width=0.8\textwidth]{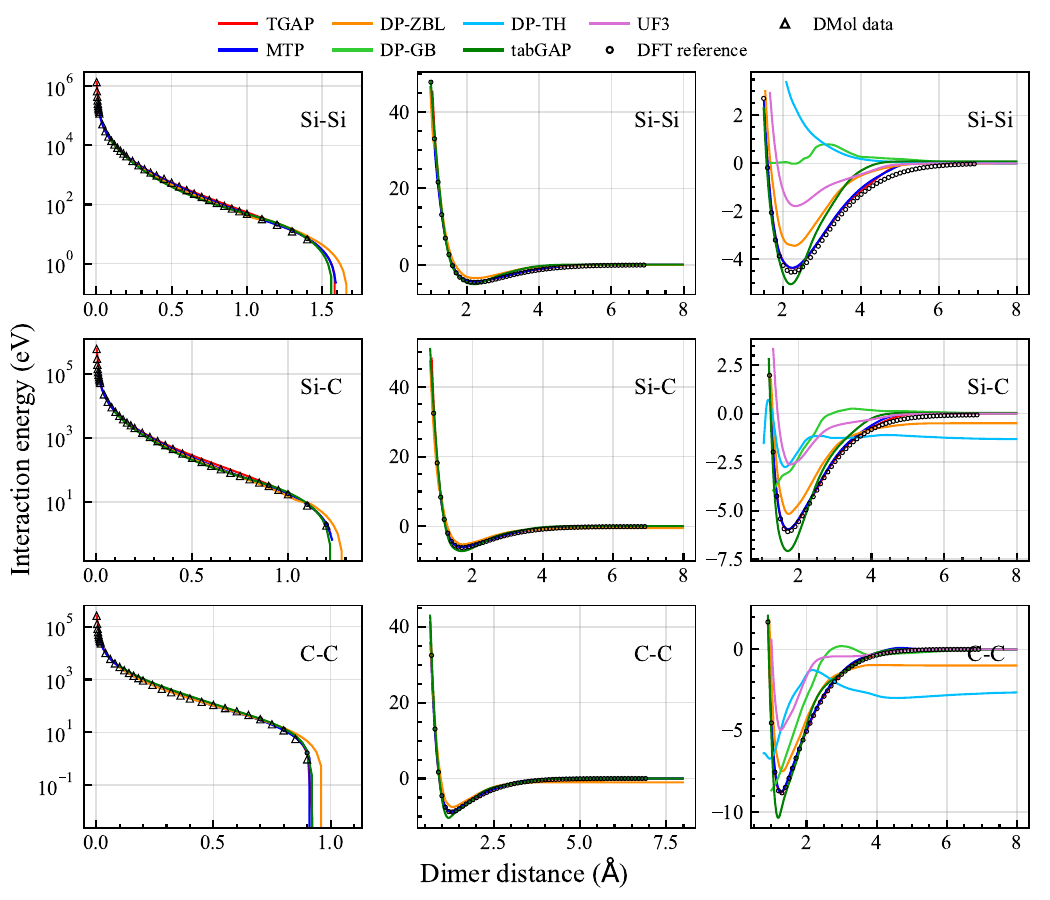}
	\caption{Interaction energies of isolated Si--Si, Si--C, and C--C dimers calculated with the MLIPs. The DFT reference curves are from the present work. The DMOL data are all-electron calculations reported in Ref.~\cite{NLH} and used to fit the NLH repulsive potential, $V_\mathrm{NLH}$. TGAP and MTP are coupled to $V_\mathrm{NLH}$ at short interatomic separations.}
	\label{fig:dimer}
\end{figure*}

Figure~\ref{fig:dimer} compares the Si--Si, Si--C, and C--C dimer interaction energies predicted by the MLIPs with our DFT calculations and, at short distances, the all-electron DMOL data used to construct $V_\mathrm{NLH}$. Results for the empirical potentials are presented in Supplementary Note~7.

The rightmost column of Figure~\ref{fig:dimer} shows the near-equilibrium and longer-range portions of the dimer interactions. TGAP and MTP closely follow the DFT reference curves over most of this range. The principal exception is the Si--Si interaction beyond approximately 5.5~\AA, where the predicted energy approaches its asymptotic value because this distance lies beyond the descriptor cutoff used by TGAP and MTP.

The middle column shows the intermediate-distance regime in which the description transitions from the MLIP contribution to the short-range repulsive interaction. For all three dimer species, TGAP and MTP exhibit a smooth connection between the DFT-like bonding regime and $V_\mathrm{NLH}$. At the shortest separations, shown in the leftmost column, the interaction is dominated by $V_\mathrm{NLH}$ and therefore follows the all-electron DMOL reference data.

Among the other MLIPs considered, DP-ZBL and tabGAP also include explicit short-range repulsive corrections and are therefore shown across the repulsive regime. The remaining models are displayed only over the near-equilibrium range. Within this regime, UF3 reproduces the dimer minima reasonably well, whereas DP-TH and DP-GB show larger deviations from the DFT reference curves. Results for FLARE-23 and FLARE-26 are discussed separately in Supplementary Note~7.

Overall, the dimer scans show that TGAP and MTP retain an accurate description of near-equilibrium bonding while transitioning smoothly to a physically constrained repulsive interaction at short distances. Dimer and compressed equation-of-state calculations have previously proven useful for constraining precisely this portion of the potential-energy surface in models intended for radiation-damage simulations~\cite{beland2017shortrange,ghorbani2025shortdistance}. This combination is essential for displacement-cascade simulations, in which energetic collisions routinely sample interatomic separations far outside the near-equilibrium training regime~\cite{beland2016extreme}.

\subsection{Threshold displacement energy}

The threshold displacement energy (TDE) is a fundamental quantity in radiation-damage modeling~\cite{r8,ghorbani2026threshold}. It is defined as the minimum kinetic energy that must be imparted to a lattice atom to produce a defect that remains stable after the collision event. The TDE therefore influences the onset and extent of primary defect production in atomistic and transport simulations and can be particularly sensitive to the short-range portion of the interatomic potential~\cite{ghorbani2025shortdistance}. Because displacement processes depend strongly on the local atomic geometry and dynamic collision pathway encountered by the recoil, the TDE is anisotropic and can vary substantially with crystallographic direction, including between closely spaced recoil directions~\cite{r8,ghorbani2026threshold}. Direction-resolved TDEs are useful for characterizing this anisotropy, whereas an angularly averaged TDE provides a complementary measure of the overall displacement response of the crystal.

We calculated TDEs along several low-index crystallographic directions for which \textit{ab initio} and experimental values are available in the literature. To characterize the broader angular dependence, we additionally calculated TDEs over a uniformly sampled set of recoil directions, from which the global average, minimum, and maximum values were determined. Simulations were performed at 300~K using a non-cubic $7\times7\times8$ supercell containing 3136 atoms. The simulation protocol and criteria used to identify stable defects are described in the Methods section.

Table~\ref{tab:tde} compares the low-index TDEs predicted by TGAP and MTP with available experimental and \textit{ab initio} values, as well as selected MLIP and empirical-potential results.

\begin{table*}[ht]
	\centering
	\caption{Threshold displacement energies (TDEs, in eV) along low-index crystallographic directions in 3C-SiC. TGAP and MTP calculations were performed at 300~K. For Si recoils, Si \hkl[111] and Si \hkl[-1-1-1] denote the close and open directions, respectively; for C recoils, the correspondence is reversed, with C \hkl[111] and C \hkl[-1-1-1] denoting the open and close directions. The simulation temperature was not specified in Ref.~\cite{tde_dft4} and is therefore assumed to be 0~K.}
	\label{tab:tde}
	\begin{ruledtabular}
		\renewcommand{\arraystretch}{1.3}
		\begin{tabular}{lcccccccccc}
			& \multirow{2}{*}{TGAP}
			& \multirow{2}{*}{MTP}
			& \multirow{2}{*}{Experiment}
			& \multicolumn{4}{c}{\textit{ab initio}}
			& \multirow{2}{*}{DP-ZBL}
			& \multirow{2}{*}{tabGAP}
			& \multirow{2}{*}{Tersoff-ZBL} \\
			\cline{5-8}
			&
			&
			&
			&
			300~K\textsuperscript{a}
			& 300~K\textsuperscript{b}
			& 100~K\textsuperscript{c}
			& 0~K\textsuperscript{d}
			&
			&
			&
			\\
			\hline

			Si \hkl[100]
			& 34.0
			& 29.5
			& 23.0\textsuperscript{e}, 25.0\textsuperscript{f}, 45.0\textsuperscript{e}
			& 41.0 & 46.0 & 49.5 & 49.5
			& 33.5 & 26.0 & 42.0 \\

			Si \hkl[110]
			& 63.5
			& 45.5
			& 23.0\textsuperscript{e}, 45.0\textsuperscript{e}
			& 50.0 & 45.0 & 70.0 & 70.0
			& 86.0 & 32.5 & 58.0 \\

			Si \hkl[111]
			& 20.5
			& 19.0
			&
			& 21.0 & 22.0 & 105.0 & 99.0
			& 20.0 & 17.5 & 21.5 \\

			Si \hkl[-1-1-1]
			& 48.5
			& 27.5
			& 18.0\textsuperscript{g}, 35.0\textsuperscript{g}, 38.0\textsuperscript{e}, 75.0\textsuperscript{e}
			& 33.0 & 21.0 & 62.0 & 63.5
			& 17.5 & 16.5 & 44.5 \\

			C \hkl[100]
			& 16.5
			& 14.0
			&
			& 18.0 & 18.0 & 20.0 & 19.0
			& 15.0 & 15.0 & 16.5 \\

			C \hkl[110]
			& 20.5
			& 18.0
			&
			& 19.0 & 14.0 & 22.5 & 21.5
			& 19.0 & 22.0 & 31.0 \\

			C \hkl[111]
			& 18.5
			& 15.5
			&
			& 17.0 & 38.0 & 20.5 & 18.0
			& 33.0 & 16.0 & 24.0 \\

			C \hkl[-1-1-1]
			& 52.5
			& 36.5
			& 22\textsuperscript{g}
			& 50.0 & 16.0 & 47.5 & 47.5
			& 14.0 & 29.5 & 53.5 \\

		\end{tabular}

		\footnotetext{Ref.~\cite{r16}, 64-atom cell}
		\footnotetext{Ref.~\cite{tde_dft2}, 64-atom cell}
		\footnotetext{Ref.~\cite{tde_dft3}, 1000-atom cell}
		\footnotetext{Ref.~\cite{tde_dft4}, 216-atom cell}
		\footnotetext{Ref.~\cite{tde-exp3}, 140--1450~K}
		\footnotetext{Refs.~\cite{tde-exp1,tde-exp4}, 300~K}
		\footnotetext{Ref.~\cite{tde-exp5}}

	\end{ruledtabular}
\end{table*}

For comparison with other interatomic models, we restrict the MLIP results in Table~\ref{tab:tde} to DP-ZBL and tabGAP, since these are the other models considered here that combine explicit short-range repulsive interactions with defective configurations in their training datasets. Tersoff-ZBL is included as an empirical reference because it gives the closest overall agreement among the classical potentials examined in our benchmark. Results for the remaining empirical potentials are provided in Supplementary Note~8.

The TDE depends strongly on both the recoil species and crystallographic direction. As noted in Table~\ref{tab:tde}, the \hkl[111] and \hkl[-1-1-1] directions are crystallographically inequivalent in zinc-blende SiC because of the two different sublattices. For Si recoils, \hkl[111] and \hkl[-1-1-1] correspond to the close and open directions, respectively, whereas the assignment is reversed for C recoils. In the open direction, a C recoil moves toward the octahedral interstitial region, with a tetrahedral interstitial site lying further along the trajectory at sufficiently high recoil energy. In the close direction, the recoil instead encounters the nearest-neighbor Si atom.

The literature values in Table~\ref{tab:tde} show substantial variation, particularly for Si recoils. The Si \hkl[-1-1-1] direction provides a striking example: experimental TDEs range from 18 to 75~eV, while the available \textit{ab initio} predictions also differ considerably. Such variation reflects both the complexity of the underlying displacement trajectory~\cite{tde_dft5} and differences among experimental techniques and irradiation conditions. Ref.~\cite{tde-exp1}, for example, reports systematically lower thresholds obtained using cathodoluminescence, photoluminescence, and electron paramagnetic resonance than those inferred from transmission electron microscopy, Rutherford backscattering spectroscopy, or positron annihilation spectroscopy. Experimental TDEs can also depend on irradiation temperature and flux, while angular dispersion of low-energy incident electrons can reduce the directional resolution of the measurement~\cite{tde-exp1}.

The spread among the \textit{ab initio} results may likewise reflect differences in simulation size and numerical convergence. Ref.~\cite{tde_dft3} employed a 1000-atom supercell and a 90~Ry basis-set cutoff, compared with supercells of 64--216 atoms and cutoffs of 35--60~Ry in the other calculations listed in Table~\ref{tab:tde}. The TGAP and MTP predictions for Si recoils are generally lower than those reported in Ref.~\cite{tde_dft3}, whereas their C-recoil TDEs are in closer agreement. A similar tendency is observed for tabGAP and Tersoff-ZBL.

Given the substantial spread in both the experimental and \textit{ab initio} references, comparison is most informative at the level of ranges and directional trends rather than against any single value. The TGAP and MTP predictions for Si \hkl[100] and Si \hkl[-1-1-1] fall within the reported experimental ranges. For Si \hkl[110], the MTP prediction of 45.5~eV lies at the upper end of the reported experimental range, while tabGAP also falls within that range; TGAP predicts a higher threshold and DP-ZBL substantially higher still. Overall, the comparison illustrates both the strong directional dependence of TDE in 3C-SiC and the uncertainty associated with establishing a unique reference value for individual recoil directions.

\begin{figure*}[htbp!]
	\centering
	
	\begin{subfigure}[t]{0.45\textwidth}
		\centering
		\includegraphics[width=\textwidth]{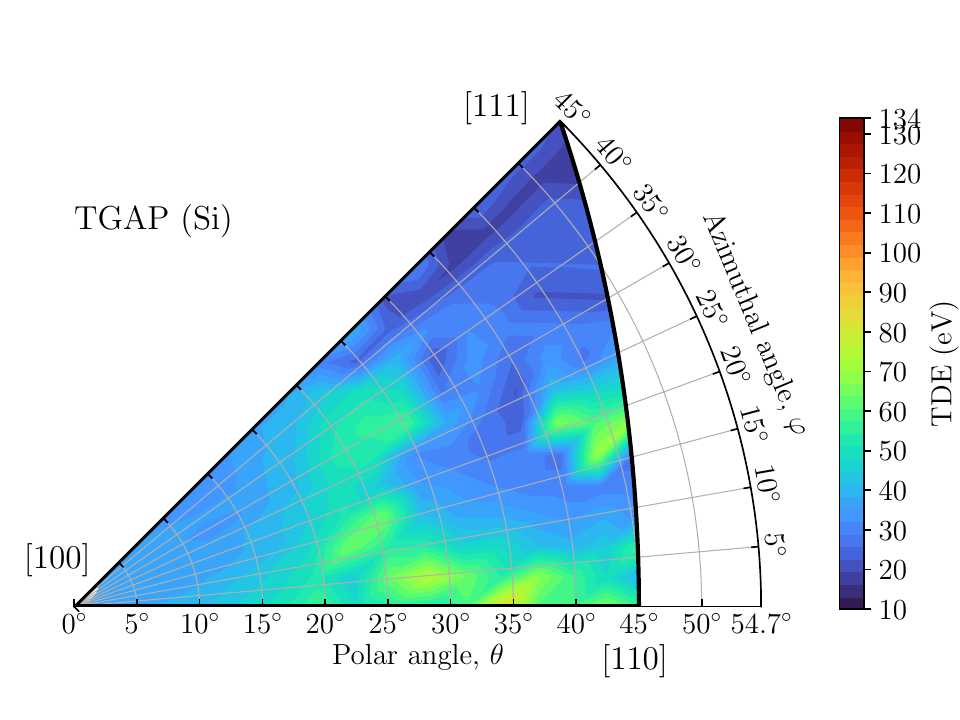}
	\end{subfigure}
	\begin{subfigure}[t]{0.45\textwidth}
		\centering
		\includegraphics[width=\textwidth]{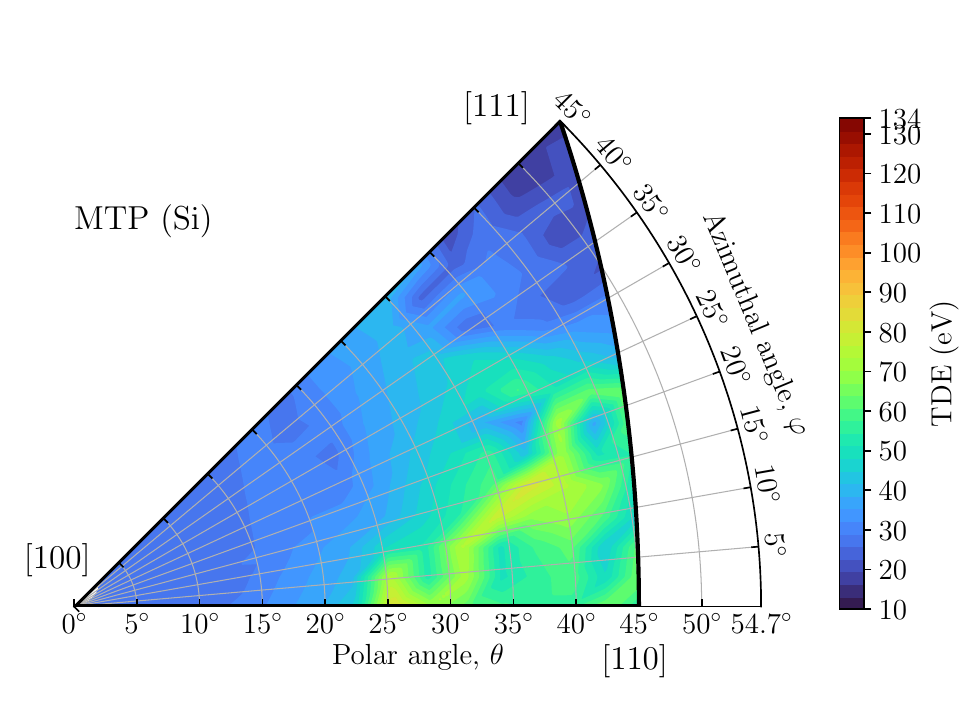}
	\end{subfigure}
	
	\vspace{0.0cm}
	
	\begin{subfigure}[t]{0.45\textwidth}
		\centering
		\includegraphics[width=\textwidth]{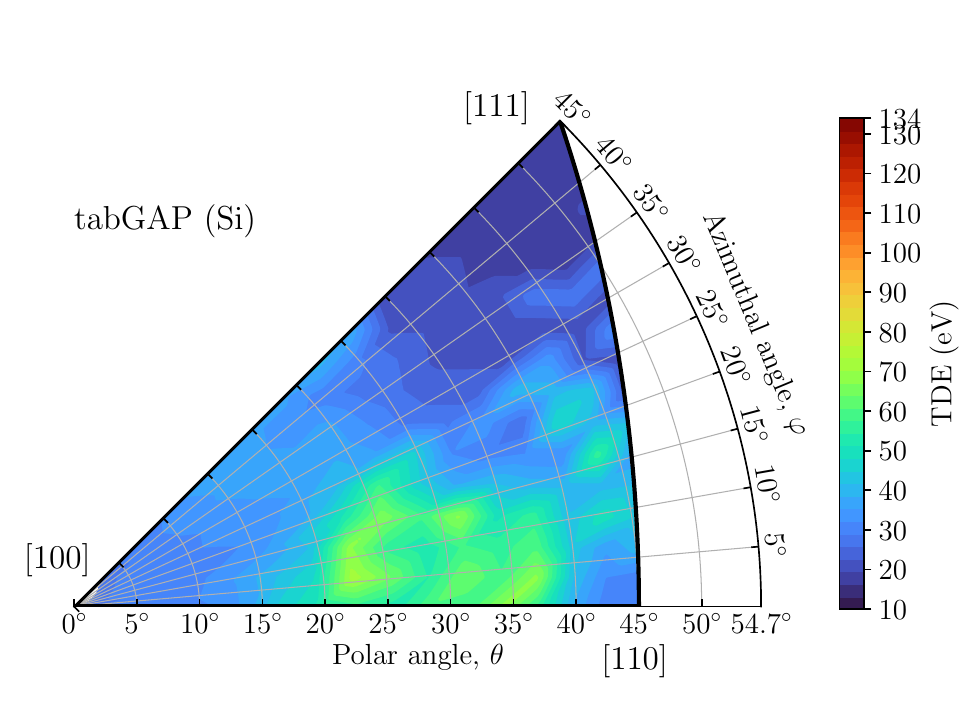}
	\end{subfigure}
	\begin{subfigure}[t]{0.45\textwidth}
		\centering
		\includegraphics[width=\textwidth]{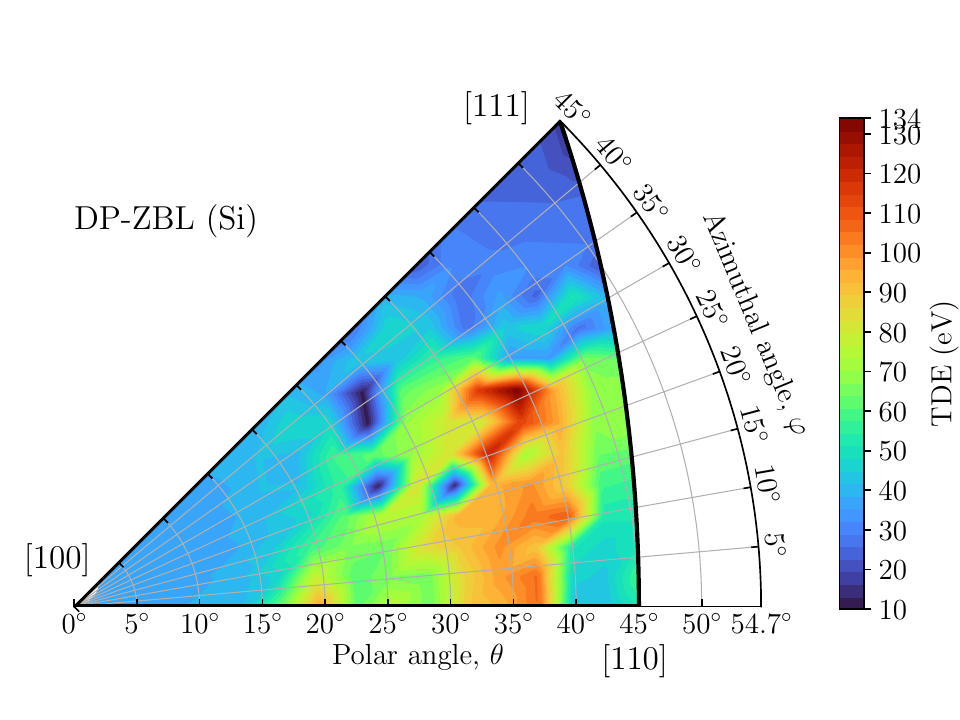}
	\end{subfigure}
	
	\caption{Directional TDE maps for Si recoils in 3C-SiC calculated using the MLIPs at 300~K. A total of 154 uniformly distributed recoil directions were sampled, with linear interpolation used between sampled directions for visualization. The global average, minimum, and maximum TDEs for each potential are reported in Table~\ref{tab:ave-tde}.}
	\label{fig:maps-Si}
\end{figure*}

Angular maps of the TDE were constructed to examine recoil directions beyond the low-index orientations listed in Table~\ref{tab:tde}. Exploiting the symmetry of the zinc-blende structure, 154 recoil directions were sampled using the uniform-sampling procedure described in Ref.~\cite{uniform-sampling}. All simulations were performed at 300~K. The resulting maps are shown in Figure~\ref{fig:maps-Si} for Si recoils and Figure~\ref{fig:maps-C} for C recoils. Corresponding results for the empirical potentials are presented in Supplementary Note~8.

\begin{figure*}[htbp!]
	\centering
	
	\begin{subfigure}[t]{0.45\textwidth}
		\centering
		\includegraphics[width=\textwidth]{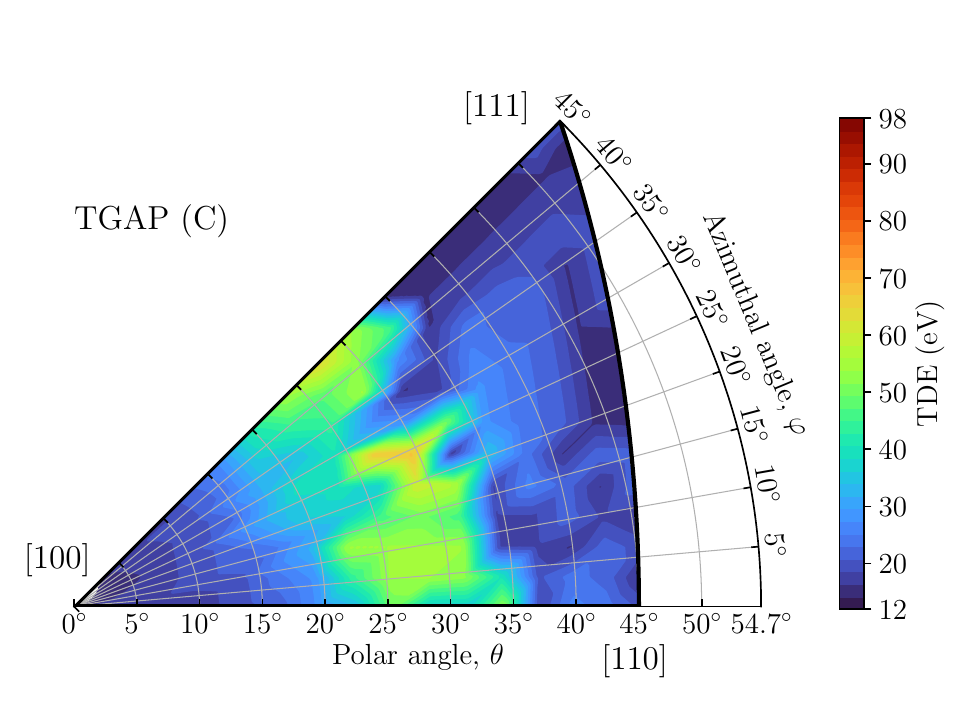}
	\end{subfigure}
	\begin{subfigure}[t]{0.45\textwidth}
		\centering
		\includegraphics[width=\textwidth]{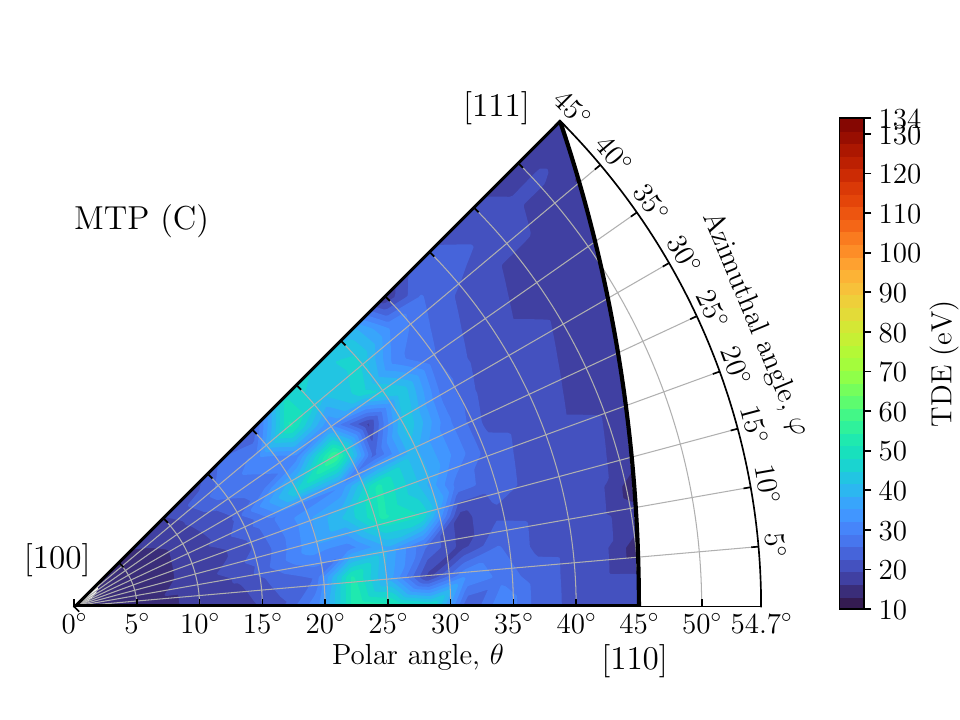}
	\end{subfigure}
	
	\vspace{0.0cm}
	
	\begin{subfigure}[t]{0.45\textwidth}
		\centering
		\includegraphics[width=\textwidth]{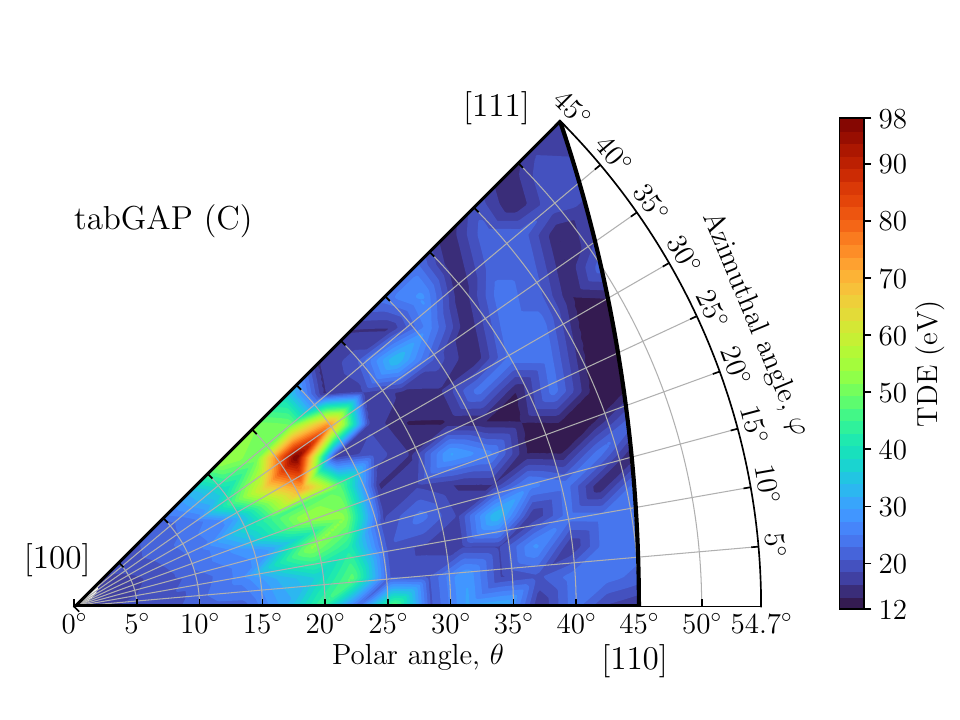}
	\end{subfigure}
	\begin{subfigure}[t]{0.45\textwidth}
		\centering
		\includegraphics[width=\textwidth]{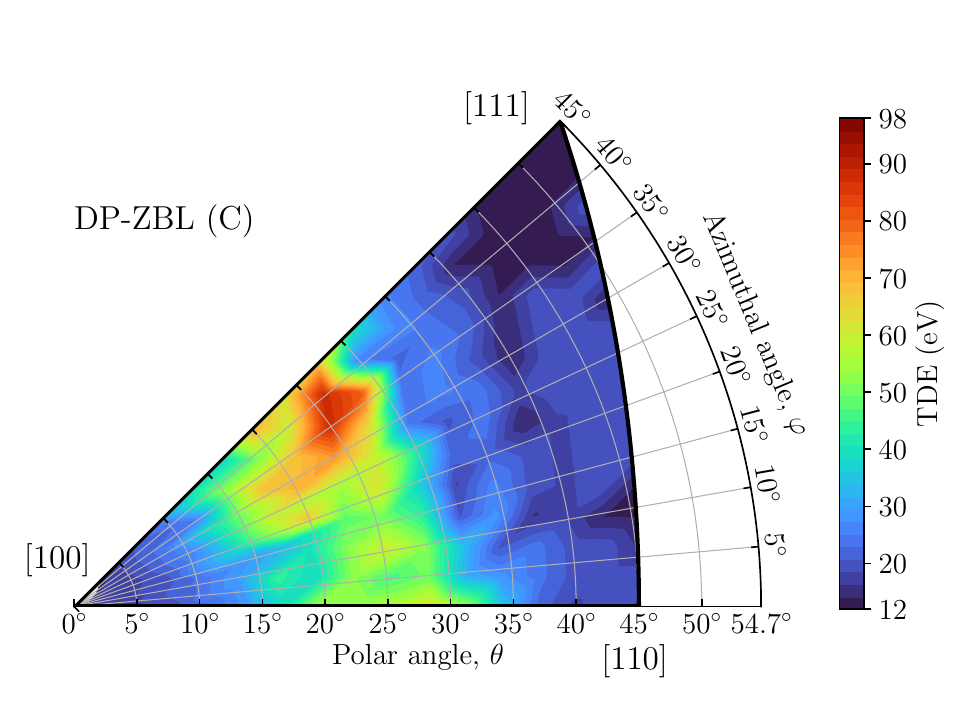}
	\end{subfigure}
	
	\caption{Directional TDE maps for C recoils in 3C-SiC calculated using the MLIPs at 300~K. A total of 154 uniformly distributed recoil directions were sampled, with linear interpolation used between sampled directions for visualization. The global average, minimum, and maximum TDEs for each potential are reported in Table~\ref{tab:ave-tde}.}
	\label{fig:maps-C}
\end{figure*}

The directional dependence predicted for Si recoils shows greater model-to-model variation than that for C recoils. As shown in Figure~\ref{fig:maps-Si}, all four MLIPs predict a region of elevated TDE within approximately $20^\circ < \theta < 40^\circ$, indicating increased resistance to stable defect formation over this angular range. The position, magnitude, and angular extent of this feature, however, differ substantially among the models. TGAP and tabGAP place the region of highest TDE near $\theta \approx 25^\circ$, whereas MTP and DP-ZBL shift it toward $\theta \approx 35^\circ$ and maintain elevated values toward the \hkl[110] boundary. DP-ZBL predicts both the largest local TDEs and the broadest high-TDE region, particularly for $\phi < 30^\circ$, followed in overall magnitude and extent by MTP, TGAP, and tabGAP. Thus, although the models reproduce a broadly similar angular pattern for Si recoils, they differ considerably in the magnitude and localization of the most resistant directions.

For C recoils, the angular distributions are more similar across the four models. Figure~\ref{fig:maps-C} shows a common band of elevated TDE concentrated primarily within $15^\circ < \theta < 30^\circ$. Differences remain in the magnitude and localization of this feature: DP-ZBL and tabGAP predict larger local maxima, whereas TGAP and MTP produce smoother and more uniform angular distributions. Overall, the C-recoil maps show greater agreement among the MLIPs in the location of the principal high-TDE regions than the Si-recoil maps, although appreciable differences in their magnitude remain.

We calculated the orientation-averaged TDE by integrating the directional TDE over solid angle according to
\begin{equation}
	TDE_\mathrm{ave} =
	\frac{\iint T(\theta,\varphi)\sin\theta\,\mathrm{d}\theta\,\mathrm{d}\varphi}
	{\iint \sin\theta\,\mathrm{d}\theta\,\mathrm{d}\varphi},
	\label{eq:ave-tde}
\end{equation}
where $T(\theta,\varphi)$ is the TDE for the recoil direction $(\theta,\varphi)$. The resulting orientation-averaged values, together with the minimum and maximum TDEs sampled for each potential, are reported in Table~\ref{tab:ave-tde}.

\begin{table}[ht]
	\centering
	\caption{Orientation-averaged, minimum, and maximum threshold displacement energies (TDEs) obtained from the directional simulations at 300~K. The orientation-averaged values are calculated using Eq.~\ref{eq:ave-tde}. All values are in eV.}
	\label{tab:ave-tde}
	\begin{ruledtabular}
		\renewcommand{\arraystretch}{1.3}
		\begin{tabular}{lcccc}
			& TGAP & MTP & tabGAP & DP-ZBL \\
			\hline
			$\mathrm{TDE}^{\mathrm{Si}}_{\mathrm{ave}}$ & 37 & 41 & 34 & 54 \\
			$\mathrm{TDE}^{\mathrm{Si}}_{\mathrm{min}}$  & 18 & 16 & 18 & 10 \\
			$\mathrm{TDE}^{\mathrm{Si}}_{\mathrm{max}}$  & 84 & 84 & 72 & 134 \\
			\hline
			$\mathrm{TDE}^{\mathrm{C}}_{\mathrm{ave}}$   & 26 & 24 & 22 & 27 \\
			$\mathrm{TDE}^{\mathrm{C}}_{\mathrm{min}}$   & 14 & 14 & 14 & 14 \\
			$\mathrm{TDE}^{\mathrm{C}}_{\mathrm{max}}$   & 68 & 58 & 98 & 88 \\
		\end{tabular}
	\end{ruledtabular}
\end{table}

For all four MLIPs, the orientation-averaged TDE is higher for Si recoils than for C recoils. The difference between $\mathrm{TDE}^{\mathrm{Si}}_{\mathrm{ave}}$ and $\mathrm{TDE}^{\mathrm{C}}_{\mathrm{ave}}$ ranges from 11~eV for TGAP to 27~eV for DP-ZBL. The Si-recoil distributions also show greater model dependence: DP-ZBL predicts both the largest orientation-averaged TDE and the broadest range of sampled values, with a minimum of 10~eV and a maximum of 134~eV. By comparison, TGAP, MTP, and tabGAP give Si-recoil minima of 16--18~eV. For C recoils, the orientation-averaged values fall within a narrower range of 22--27~eV, and all four models predict the same sampled minimum of 14~eV, although their maximum values differ substantially.

The agreement among TGAP, MTP, and tabGAP therefore suggests a characteristic minimum directional threshold of approximately $17\pm1$~eV for Si recoils, while all four models give a minimum of 14~eV for C recoils. These values should be distinguished from the orientation-averaged TDEs, which are substantially higher and characterize a different aspect of the displacement response. The appropriate quantity for use in reduced radiation-damage or transport models depends on how directional anisotropy is represented; accordingly, the minimum and orientation-averaged values are both reported here to facilitate comparison and model selection.

\subsection{Defects}

The training dataset contains a broad set of intrinsic point defects and small defect complexes, including vacancies, antisites, split interstitials, tetrahedral and hexagonal interstitials, divacancies, and trivacancies. Bond-centered interstitials were not included because the present dataset focuses on intrinsic defects, whereas bond-centered configurations are more commonly relevant to extrinsic interstitial species~\cite{bc1,defs-5}. For split interstitials, eight configurations were constructed by considering both Si and C interstitial species on the Si and C sublattices, with \hkl[100] and \hkl[110] orientations. We denote Si$^\mathrm{i}$C and C$^\mathrm{i}$C as split interstitial pairs centered on a C lattice site, and Si$^\mathrm{i}$Si and C$^\mathrm{i}$Si as the corresponding configurations centered on a Si lattice site. Four tetrahedral interstitial configurations were included, distinguished by the interstitial species and the surrounding tetrahedral cage; for example, Si$_\mathrm{TC}$ denotes a Si interstitial surrounded by four C atoms. Two hexagonal interstitials were also considered, distinguished by the interstitial species, with C$_\mathrm{H}$ denoting the C hexagonal interstitial. The vacancy-containing configurations comprise the two monovacancies, one divacancy, and two trivacancies with Si--C--Si and C--Si--C local compositions. Further details of the defect structures are provided in Supplementary Note~1.

The defect formation energies were calculated as
\begin{equation}
	\label{eq:ef}
	E_\mathrm{f}
	=
	E_\mathrm{defective}
	-
	E_\mathrm{pristine}
	+
	n_{\mathrm{Si}}\mu_{\mathrm{Si}}
	+
	n_{\mathrm{C}}\mu_{\mathrm{C}},
\end{equation}
where $E_\mathrm{pristine}$ and $E_\mathrm{defective}$ are the total energies of the pristine and defective supercells, respectively. The quantity $n_\mathrm{X}$ is defined as
$n_\mathrm{X}=n_\mathrm{X}^{\mathrm{pristine}}-n_\mathrm{X}^{\mathrm{defective}}$,
such that atoms removed from the cell contribute positively and added atoms negatively in Eq.~\ref{eq:ef}. The elemental chemical potentials are constrained by equilibrium with 3C-SiC. Formation energies were evaluated in the Si-rich limit, with
$\mu_\mathrm{Si}=\mu_\mathrm{Si}^{\mathrm{diamond}}$
and
$\mu_\mathrm{C}=\mu_\mathrm{SiC}-\mu_\mathrm{Si}$,
where $\mu_\mathrm{SiC}$ is the chemical potential per SiC formula unit in cubic 3C-SiC. Details of the corresponding DFT calculations are provided in Supplementary Note~4.

Table~\ref{tab:efs} compares the neutral-defect formation energies predicted by TGAP and MTP with our DFT calculations, literature DFT values, and selected existing MLIPs. Results obtained with empirical potentials are presented in Supplementary Note~9. For configurations that are unstable upon relaxation, the final relaxed structure is reported together with its formation energy where applicable.

\begin{table*}[ht]
	\centering
	\caption{Defect formation energies, $E_\mathrm{f}$ (eV), in 3C-SiC calculated with the MLIPs and DFT. Literature DFT values are included where available. Formation energies were evaluated in the Si-rich limit. V$_\mathrm{X}$ denotes a vacancy on sublattice X; Si$_\mathrm{C}$ denotes a Si antisite on the C sublattice; D$_\mathrm{vac}$ denotes the divacancy; and T$_\mathrm{vac}^{\mathrm{X}}$ denotes a trivacancy centered on species X. X$_\mathrm{TY}$, X$_\mathrm{H}$, and X$^\mathrm{i}$Y$_{\hkl[hkl]}$ denote tetrahedral, hexagonal, and split interstitial configurations, respectively. For configurations that are unstable upon relaxation, the resulting defect geometry and, where applicable, its formation energy are reported.}
	\label{tab:efs}
	\begin{ruledtabular}
		\renewcommand{\arraystretch}{1.3}
		\begin{tabular}{lcccccc}
			& TGAP & MTP & tabGAP & DP-ZBL\textsuperscript{a} & DFT & DFT$_{\mathrm{lit}}$ \\
			\hline
			V$_{\mathrm{Si}}$
			& 7.28 & 6.47 & 2.23 & 5.66 & 7.72 & 8.40\textsuperscript{c} \\

			V$_{\mathrm{C}}$
			& 3.77 & 3.05 & 1.09 & 5.12 & 4.20 & 4.23\textsuperscript{c} \\

			Si$_\mathrm{TSi}$
			& Si$_{\mathrm{HC}}$(9.76) & Si$_\mathrm{TSi}$(8.51) & Si$_\mathrm{TSi}$(10.14) & Si$_{\mathrm{HC}}$(10.93) & Si$_{\mathrm{HC}}$(10.21) & 10.20\textsuperscript{b} \\

			Si$_\mathrm{TC}$
			& 8.02 & 6.19 & 6.99 & 8.80 & 8.46 & 8.63\textsuperscript{b} \\

			C$_\mathrm{TC}$
			& 10.34 & 8.99 & 12.02 & 8.89 & 10.96 & 10.42\textsuperscript{b} \\

			C$_\mathrm{TSi}$
			& 9.51 & 7.55 & 9.72 & 7.57 & 9.96 & 9.67\textsuperscript{b} \\

			Si$_\mathrm{C}$
			& 3.03 & 3.15 & 1.54 & 6.66 & 3.33 & 3.37\textsuperscript{c} \\

			C$_\mathrm{Si}$
			& 3.70 & 3.62 & 4.95 & 0.66 & 3.94 & 4.16\textsuperscript{c} \\

			Si$^\mathrm{i}$C$_{\hkl[100]}$
			& 9.13 & 8.42 & 7.23 & 10.63 & 10.20 & 9.98\textsuperscript{b} \\

			Si$^\mathrm{i}$Si$_{\hkl[100]}$
			& 8.72 & 7.80 & 6.45 & 9.60 & 9.54 & 9.62\textsuperscript{b} \\

			Si$^\mathrm{i}$Si$_{\hkl[110]}$
			& 8.25 & 7.91 & 8.29 & 10.18 & 8.37 & 8.53\textsuperscript{b} \\

			Si$^\mathrm{i}$C$_{\hkl[110]}$
			& Si$_\mathrm{TC}$
			& Si$_\mathrm{TC}$
			& Si$^\mathrm{i}$C$_{\hkl[331]}$ (6.84)
			& Si$_\mathrm{TC}$
			& Si$_\mathrm{TC}$
			& Si$_\mathrm{TC}$\textsuperscript{b} \\

			C$^\mathrm{i}$C$_{\hkl[100]}$
			& 5.92 & 4.79 & 6.37 & 4.74 & 6.79 & 6.58\textsuperscript{b} \\

			C$^\mathrm{i}$Si$_{\hkl[100]}$
			& 6.67 & 5.77 & 6.49 & 6.52 & 7.60 & 7.27\textsuperscript{b} \\

			C$^\mathrm{i}$C$_{\hkl[110]}$
			& 6.49 & 5.37 & 6.69 & 5.72 & 7.14 & 6.89\textsuperscript{b} \\

			C$^\mathrm{i}$Si$_{\hkl[110]}$
			& CC$_{\hkl[001]}$
			& CC$_{\hkl[001]}$
			& C$^\mathrm{i}$Si$_{\hkl[110]}$ (9.03)
			& CC$_{\hkl[001]}$
			& CC$_{\hkl[225]}$ (6.72)
			& CC$_{\hkl[225]}$ (6.41)\textsuperscript{b} \\

			Si$_{\mathrm{H}}$
			& Si$_\mathrm{TC}$
			& Si$_\mathrm{TC}$
			& Si$_\mathrm{TC}^*$
			& Si$_\mathrm{TC}$
			& Si$_\mathrm{TC}$
			& \\

			C$_{\mathrm{H}}$
			& C$_{\mathrm{H}}^{\mathrm{C-close}}$ (8.72)
			& C$_{\mathrm{H}}^{\mathrm{C-close}}$ (7.21)
			& C$_{\mathrm{H}}^{\mathrm{Si-close}}$ (9.79)
			& CC$_{\hkl[001]}$
			& C$_{\mathrm{H}}^{\mathrm{C-close}}$ (9.27)
			& 8.21\textsuperscript{d} \\

			D$_\mathrm{vac}$
			& 7.61 & 7.20 & 2.44 & 8.22 & 7.33 & \\

			T$_{\mathrm{vac}}^{\mathrm{Si}}$
			& 8.17 & 7.57 & -0.30 & 10.76 & 7.70 & \\

			T$_{\mathrm{vac}}^{\mathrm{C}}$
			& 12.26 & 12.05 & 6.79 & 11.76 & 13.10 & \\
		\end{tabular}

		\footnotetext{Ref.~\cite{dp-zbl}}
		\footnotetext{Ref.~\cite{defs-rotation}}
		\footnotetext{Ref.~\cite{defs-Ag}}
		\footnotetext{Ref.~\cite{defs-hex}}
	\end{ruledtabular}
\end{table*}

TGAP generally follows the DFT formation energies closely across the defect set and reproduces their relative energetic ordering well. MTP captures the same broad trends but tends to underestimate the formation energies of the higher-energy interstitial configurations. Larger and less systematic deviations are observed for tabGAP and DP-ZBL. For example, tabGAP substantially underestimates the vacancy and vacancy-cluster formation energies and predicts a negative formation energy for T$_{\mathrm{vac}}^{\mathrm{Si}}$, indicating pronounced over-stabilization of this trivacancy configuration. DP-ZBL gives reasonable energies for several interstitials but strongly misrepresents the antisite energetics, predicting Si$_\mathrm{C}$ too high and C$_\mathrm{Si}$ too low relative to DFT.

The relaxation pathways provide an additional test beyond the absolute formation energies. In our DFT calculations, Si$_\mathrm{H}$ and Si$^\mathrm{i}$C$_{\hkl[110]}$ are unstable and both relax to Si$_\mathrm{TC}$, consistent with previous DFT studies~\cite{defs-rotation,defs-Ag,defs-4,defs-5}. TGAP, MTP, and DP-ZBL reproduce both transformations. In tabGAP, Si$^\mathrm{i}$C$_{\hkl[110]}$ instead relaxes to a Si$^\mathrm{i}$C$_{\hkl[331]}$ dumbbell, while Si$_\mathrm{H}$ relaxes to a distorted Si$_\mathrm{TC}^{*}$ configuration in which one bond is elongated relative to the other three.

The C$^\mathrm{i}$Si$_{\hkl[110]}$ configuration is likewise unstable in our DFT calculations and relaxes to a C--C dumbbell oriented close to \hkl[225], in agreement with Ref.~\cite{defs-rotation}. Other DFT studies also report instability of the initial C$^\mathrm{i}$Si$_{\hkl[110]}$ configuration, although the final orientation varies: relaxation to a tilted C$^\mathrm{i}$C$_{\hkl[100]}$ configuration was reported in Ref.~\cite{defs-hex}, whereas Ref.~\cite{defs-6} found a tilted C$^\mathrm{i}$Si$_{\hkl[110]}$ geometry. TGAP, MTP, and DP-ZBL reproduce the instability of the initial configuration but relax instead to a C--C dumbbell oriented along \hkl[001]. In tabGAP, the original C$^\mathrm{i}$Si$_{\hkl[110]}$ configuration remains locally stable.

The Si$_\mathrm{TSi}$ interstitial also relaxes in DFT, transforming to Si$_\mathrm{HC}$, corresponding to a Si interstitial in a hexagonal C cage. TGAP and DP-ZBL reproduce this transformation, whereas MTP and tabGAP retain the initial Si$_\mathrm{TSi}$ configuration. Among the Si interstitials considered, DFT identifies Si$^\mathrm{i}$Si$_{\hkl[110]}$ as the lowest-energy configuration, although its formation energy differs from that of Si$_\mathrm{TC}$ by only 0.09~eV. Reproducing this near-degenerate ordering is challenging for the tested MLIPs. For the C interstitial family, the lowest-energy relaxed structure in our DFT calculations originates from the C$^\mathrm{i}$Si$_{\hkl[110]}$ configuration; TGAP, MTP, and DP-ZBL reproduce its instability and yield comparably low-energy C--C dumbbell configurations, although with a different final orientation.

Overall, TGAP provides the closest agreement with DFT across the defect formation energies while also reproducing most of the observed relaxation pathways. MTP reproduces many of the same structural transformations and relative trends, but with a systematic tendency to lower the energies of higher-energy defects. The comparison therefore highlights both energetic accuracy and relaxation behavior as complementary tests of defect descriptions in MLIPs.

\begin{figure*}[ht]
	\centering
	\includegraphics[width=1.0\textwidth]{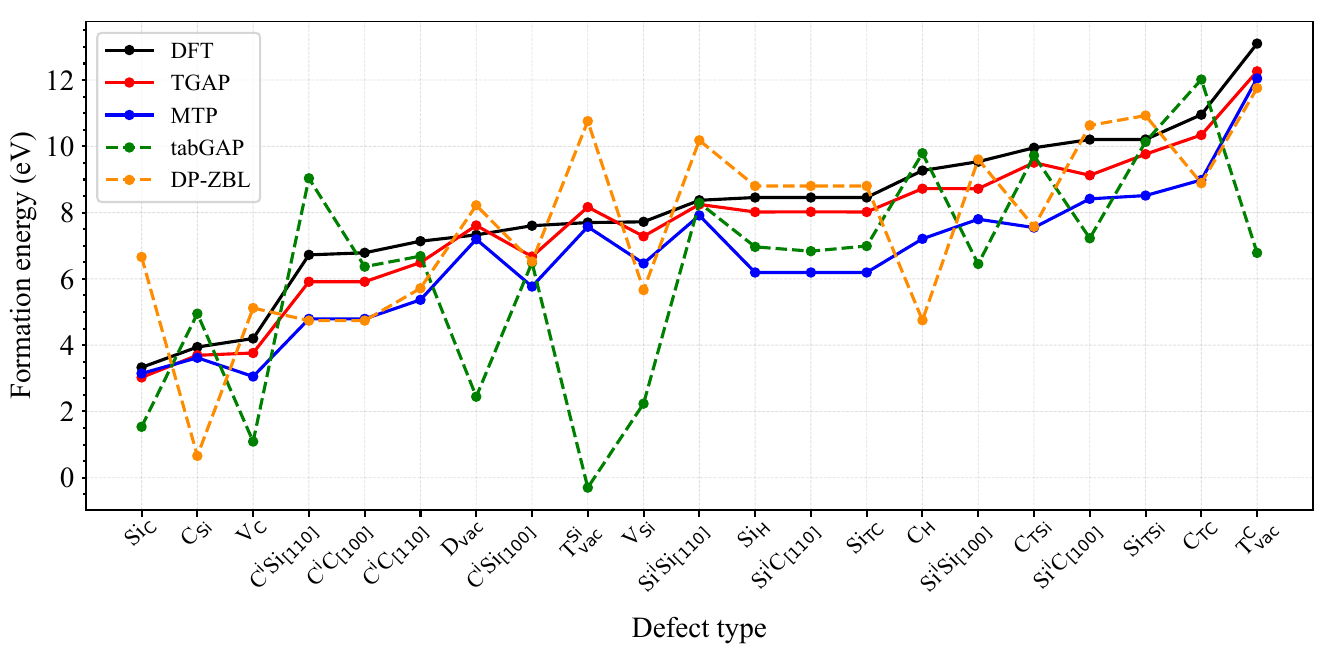}
	\caption{Comparison of the relative stability of defects in 3C-SiC predicted by the MLIPs and DFT. Defect configurations are ordered by increasing DFT formation energy, and the corresponding MLIP formation energies are plotted using the same ordering. A monotonically increasing MLIP curve therefore indicates preservation of the DFT-predicted rank ordering, whereas local decreases indicate inversions in the predicted relative stability.}
	\label{fig:sia-stability}
\end{figure*}

Figure~\ref{fig:sia-stability} provides a complementary comparison of the relative defect stabilities predicted by the MLIPs. The defect configurations listed in Table~\ref{tab:efs} are ordered according to increasing DFT formation energy, and the corresponding MLIP formation energies are plotted using this fixed DFT-based ordering.

TGAP most closely follows the DFT reference, reproducing both the magnitude of the formation energies and their relative ordering across most of the defect configurations. MTP generally preserves the overall DFT stability sequence but systematically underestimates the formation energies of intermediate- and high-energy defects, typically by approximately 1--2~eV. Its comparatively smooth increase along the DFT-based ordering nevertheless indicates that the relative energetic trends are retained more accurately than the absolute energies.

DP-ZBL and tabGAP show more frequent departures from the DFT-based stability sequence. For DP-ZBL, several configurations that are adjacent or closely spaced in DFT exhibit substantial changes in relative energy, producing local inversions in the ordering. tabGAP shows still stronger variations for several defect classes, including the pronounced stabilization of vacancy-containing configurations discussed above. Thus, although both models reproduce selected individual formation energies, their relative stability rankings are less consistent with DFT across the full defect set.

Overall, TGAP provides the closest agreement with DFT in both absolute defect formation energies and relative stability ordering. MTP reproduces the broader energetic hierarchy but with a systematic downward shift for higher-energy defects, whereas DP-ZBL and tabGAP exhibit larger configuration-dependent deviations and more frequent changes in the DFT-based rank ordering.

Vacancy migration in 3C-SiC can proceed through nearest-neighbor (NN) or second-nearest-neighbor (2NN) atomic hops~\cite{defs-4}. A 2NN hop transfers a vacancy between sites on the same sublattice and therefore corresponds directly to
$V_\mathrm{Si}\rightarrow V_\mathrm{Si}$ or
$V_\mathrm{C}\rightarrow V_\mathrm{C}$ migration. In contrast, an NN hop transfers an atom from the opposite sublattice into the vacancy and creates a vacancy--antisite complex. For example, migration of a neighboring Si atom into a C vacancy produces a $V_\mathrm{Si}\text{-}\mathrm{Si}_\mathrm{C}$ complex, whereas the analogous hop into a Si vacancy produces $V_\mathrm{C}\text{-}\mathrm{C}_\mathrm{Si}$. Previous DFT calculations indicate that the $V_\mathrm{Si}\text{-}\mathrm{Si}_\mathrm{C}$ complex is unstable, such that carbon-vacancy migration proceeds predominantly through 2NN hops, while the $V_\mathrm{C}\text{-}\mathrm{C}_\mathrm{Si}$ complex associated with silicon-vacancy migration is metastable~\cite{defs-4}.

\begin{figure}[htp!]
	\centering

	\begin{subfigure}{\columnwidth}
		\centering
		\includegraphics[width=0.9\linewidth]{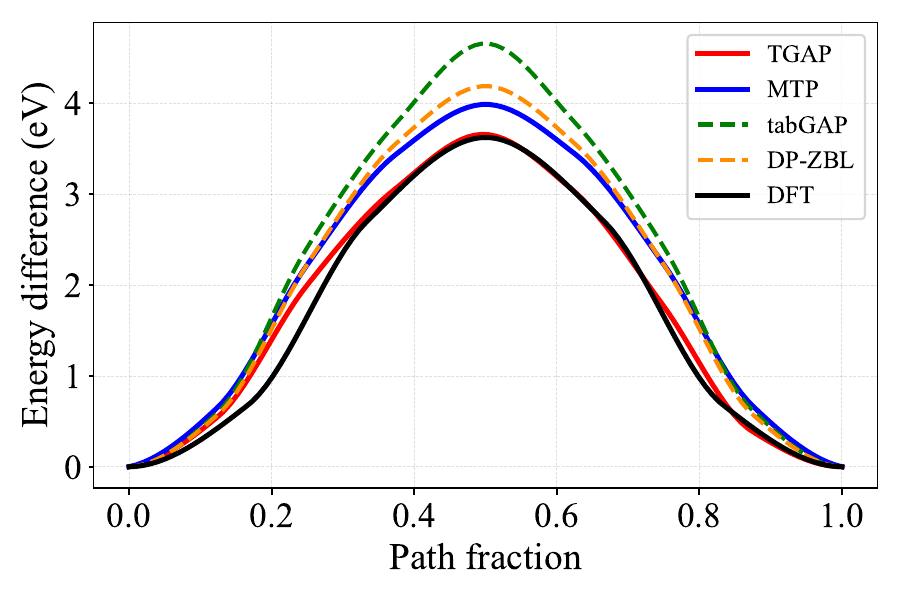}
		\caption{Second-nearest-neighbor migration of $V_\mathrm{C}$ ($V_\mathrm{C}\rightarrow V_\mathrm{C}$).}
	\end{subfigure}\\[1ex]

	\begin{subfigure}{\columnwidth}
		\centering
		\includegraphics[width=0.9\linewidth]{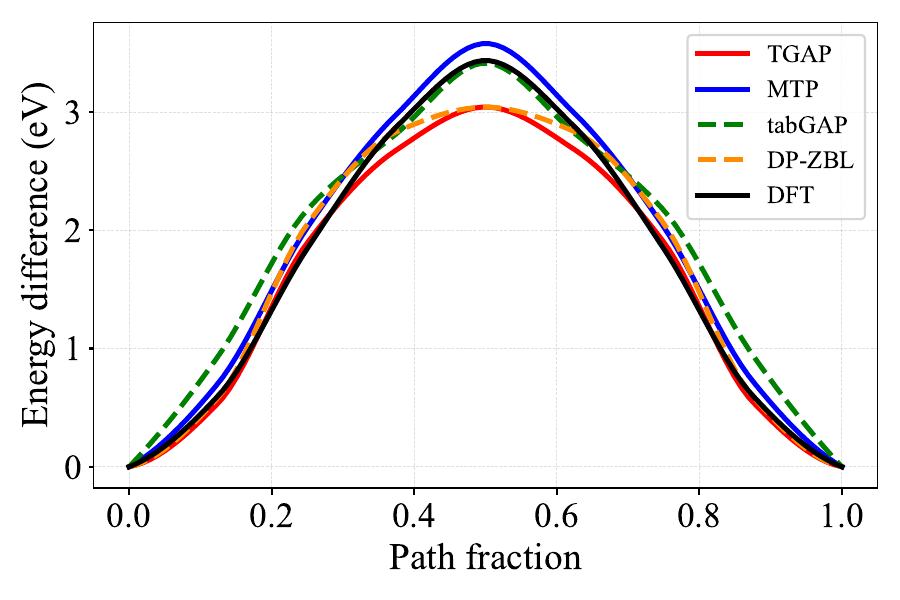}
		\caption{Second-nearest-neighbor migration of $V_\mathrm{Si}$ ($V_\mathrm{Si}\rightarrow V_\mathrm{Si}$).}
	\end{subfigure}\\[1ex]

	\begin{subfigure}{\columnwidth}
		\centering
		\includegraphics[width=0.9\linewidth]{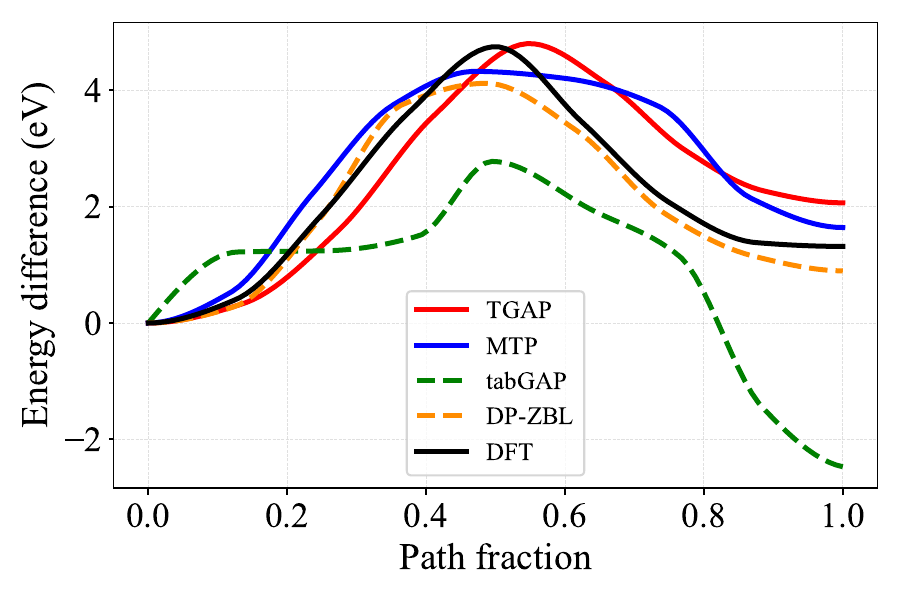}
		\caption{$V_\mathrm{Si}\text{-}\mathrm{Si}_\mathrm{C}\rightarrow V_\mathrm{Si}$ vacancy--antisite pathway associated with nearest-neighbor migration.}
	\end{subfigure}

	\caption{NEB minimum-energy paths for vacancy migration in 3C-SiC calculated with the MLIPs and DFT. Seven intermediate images were used between the endpoint configurations. Computational details are provided in Supplementary Note~4.}
	\label{fig:mep}
\end{figure}

We calculated vacancy migration pathways using the nudged elastic band (NEB) method~\cite{neb1,neb2}, as described in Supplementary Note~4. The resulting migration barriers, $E_\mathrm{m}$, are summarized in Table~\ref{tab:Em}, and the corresponding minimum-energy paths (MEPs) are compared with DFT in Figure~\ref{fig:mep}.

\begin{table*}[ht]
	\centering
	\caption{Vacancy migration barriers, $E_\mathrm{m}$ (eV), calculated with the MLIPs and DFT. Literature DFT values are included for comparison. The $V_\mathrm{Si}\rightarrow V_\mathrm{Si}$ and $V_\mathrm{C}\rightarrow V_\mathrm{C}$ pathways correspond to second-nearest-neighbor hops between sites on the same sublattice. The $V_\mathrm{Si}\text{-}\mathrm{Si}_\mathrm{C}\rightarrow V_\mathrm{Si}$ pathway probes the vacancy--antisite configuration associated with nearest-neighbor carbon-vacancy motion.}
	\label{tab:Em}
	\begin{ruledtabular}
		\renewcommand{\arraystretch}{1.3}
		\begin{tabular}{lcccccc}
			& TGAP & MTP & DP-ZBL & tabGAP & DFT & DFT$_\mathrm{lit}$ \\
			\hline
			$V_\mathrm{Si}\rightarrow V_\mathrm{Si}$
			& 3.02 & 3.58 & 3.04 & 3.41 & 3.44
			& 3.51\textsuperscript{a}, 3.4\textsuperscript{b} \\

			$V_\mathrm{C}\rightarrow V_\mathrm{C}$
			& 3.63 & 3.99 & 4.19 & 4.66 & 3.63
			& 3.64\textsuperscript{a}, 3.5\textsuperscript{b}, 3.66\textsuperscript{c} \\

			$V_\mathrm{Si}\text{-}\mathrm{Si}_\mathrm{C}\rightarrow V_\mathrm{Si}$
			& 2.74 & -- & 3.22 & -- & 3.43
			& 2.41\textsuperscript{a}, 3.5\textsuperscript{b}
		\end{tabular}

		\footnotetext{Ref.~\cite{r18}}
		\footnotetext{Ref.~\cite{defs-4}}
		\footnotetext{Ref.~\cite{Vc_2NN}}
	\end{ruledtabular}
\end{table*}

Corresponding MEPs obtained with the classical potentials are presented in Supplementary Note~9. For the $V_\mathrm{C}\rightarrow V_\mathrm{C}$ 2NN hop, TGAP reproduces the DFT migration barrier of 3.63~eV essentially exactly. MTP and DP-ZBL overestimate the barrier by 0.36 and 0.56~eV, respectively, while tabGAP gives the largest value, 1.03~eV above DFT. The MEPs show the same trend: TGAP follows the DFT energy profile most closely, whereas the other models predict progressively higher saddle-point energies.

For the $V_\mathrm{Si}\rightarrow V_\mathrm{Si}$ 2NN hop, tabGAP and MTP give the closest barriers to DFT, differing by only 0.03 and 0.14~eV, respectively. TGAP and DP-ZBL underestimate the DFT barrier by approximately 0.4~eV. Nevertheless, all four MLIPs reproduce the overall single-barrier character of the DFT pathway, with differences arising primarily in the barrier height and the detailed energy profile along the reaction coordinate.

The $V_\mathrm{Si}\text{-}\mathrm{Si}_\mathrm{C}\rightarrow V_\mathrm{Si}$ pathway provides a more demanding test. This vacancy--antisite configuration is not represented explicitly in the training dataset and therefore probes behavior outside the defect configurations used for fitting. Such tests are particularly informative for assessing generalization: in our previous Si/O/SiO$_2$ MTP, for example, a Si-vacancy migration profile was reproduced without including the corresponding NEB or vacancy-AIMD configurations in the training set~\cite{zongo2024unified}. In the present SiC benchmark, all of the MLIPs show larger deviations from the DFT MEP for this pathway than for the two 2NN vacancy hops. TGAP predicts a barrier of 2.74~eV, underestimating the present DFT value by 0.69~eV, while DP-ZBL gives the closest barrier at 3.22~eV, 0.21~eV below DFT. TGAP retains the qualitative single-barrier topology of the DFT pathway and does not introduce spurious intermediate minima, although the saddle-point region and relative energy of the endpoint configuration are not reproduced quantitatively. DP-ZBL provides the closest overall description of this pathway among the tested MLIPs.

For MTP, the energy profile develops a broad, nearly flat saddle region rather than a clearly resolved maximum, so a unique migration barrier is not assigned. tabGAP also fails to reproduce the DFT topology of the pathway: the nominal final state falls below the initial vacancy--antisite configuration, and a corresponding migration barrier is therefore not reported.

Overall, the vacancy-migration benchmark reveals a more property-dependent picture than the defect formation energies alone. TGAP most accurately reproduces the $V_\mathrm{C}$ 2NN barrier, MTP and tabGAP perform best for the $V_\mathrm{Si}$ 2NN hop, and DP-ZBL gives the closest description of the more challenging $V_\mathrm{Si}\text{-}\mathrm{Si}_\mathrm{C}\rightarrow V_\mathrm{Si}$ pathway. No single potential is uniformly most accurate across all three migration processes, and the larger discrepancies for the vacancy--antisite pathway highlight the increased difficulty of configurations not explicitly represented in the training data.

\subsection{Disordered phase}

\begin{figure*}[htp!]
	\centering
	\includegraphics[width=0.9\textwidth]{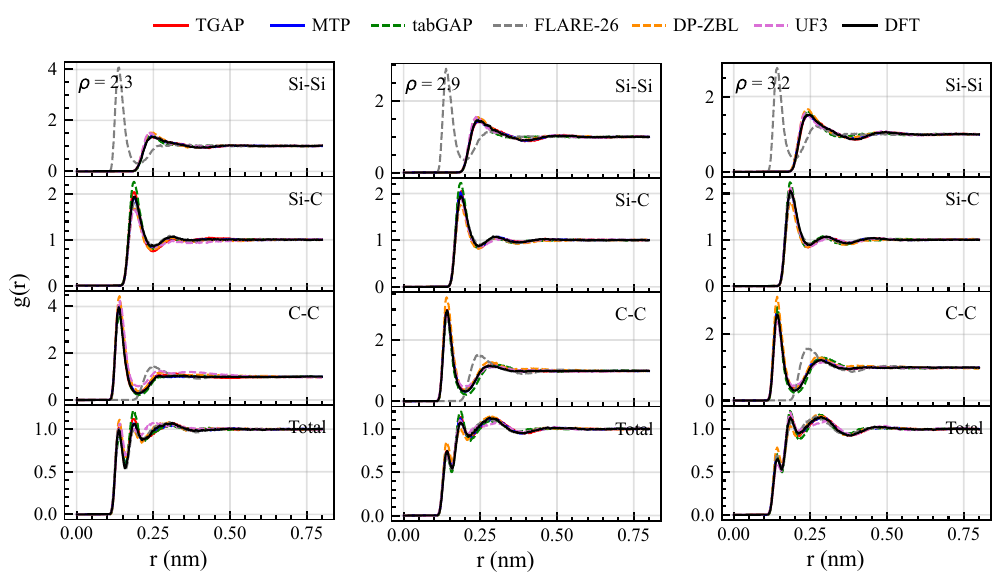}
	\caption{Radial distribution functions of liquid SiC at 5000~K calculated with the MLIPs and DFT. Results are shown for three of the six densities sampled in the liquid dataset, with densities given in g/cm$^3$. The RDFs were accumulated over 5000 trajectory frames from 216-atom cells. Results for the remaining densities are presented in Supplementary Figure~4.}
	\label{fig:rdfs}
\end{figure*}

The construction of the liquid and amorphous configurations used in the training dataset is described in Supplementary Note~1. To assess the description of the liquid structure in 3C-SiC, we calculated radial distribution functions (RDFs) at 5000~K and compared the MLIP predictions with AIMD reference data. Figure~\ref{fig:rdfs} shows results at three representative densities spanning the low, intermediate, and high portions of the sampled density range; results for the remaining densities are provided in Supplementary Note~10.

For the AIMD reference calculations, 216-atom cells at fixed density were equilibrated at 5000~K in the canonical (NVT) ensemble for 10~ps using a 1~fs timestep, followed by an additional 5~ps trajectory from which 5000 frames were used to calculate the RDFs. For the MLIP simulations, 216-atom cells at the same densities were first randomized at 6000~K for 10~ps and subsequently equilibrated at 5000~K for 15~ps in the NVT ensemble, also using a 1~fs timestep. The final 5000 frames were used for the RDF analysis.

FLARE-23, DP-GB, and DP-TH are not included in this comparison because their training datasets do not contain disordered configurations. TGAP and MTP closely reproduce the DFT liquid structure across the densities examined, including the positions and relative shapes of the principal Si--C, Si--Si, and C--C correlation peaks and the total RDF. The agreement is maintained across the sampled density range, indicating that both potentials accurately describe the liquid configurations represented by this benchmark. DP-ZBL, tabGAP, and UF3 also reproduce the DFT RDFs reasonably closely. FLARE-26 shows more pronounced deviations: a Si--Si feature develops near 0.13~nm, substantially below the nearest-neighbor separation observed in the DFT liquid, while the first C--C peak present in DFT is strongly suppressed.

Structural agreement over finite trajectories does not, however, establish numerical robustness over the much longer simulations required for large-scale applications. During extended melt--quench simulations with MTP, we occasionally observed numerical instabilities that ultimately resulted in atom loss and termination of the simulation. No such instabilities were observed with TGAP in comparable simulations. We therefore performed a separate series of constant-volume liquid simulations at 5000~K to characterize the frequency of these events as a function of density. At each density, ten independent 1728-atom simulations were run for 100~ps using a 1~fs timestep, corresponding to approximately $1.7\times10^{9}$ atom-steps of sampling per density.

No failures were observed at densities up to and including 2.6~g/cm$^3$, corresponding to pressures of approximately 10~GPa or below. At higher densities, the observed fraction of trajectories that failed increased: 2 of 10 trajectories failed at 2.75~g/cm$^3$ ($\sim14$~GPa), 2 of 10 at 2.9~g/cm$^3$ ($\sim18$~GPa), 5 of 10 at 3.05~g/cm$^3$ ($\sim22$~GPa), and all 10 trajectories failed at 3.2~g/cm$^3$. Thus, the instability becomes increasingly probable as the liquid is compressed toward the upper end of the density range represented in the MTP training data.

The failing trajectories exhibit a characteristic sequence. The instantaneous temperature first rises above the thermostat setpoint over a period of several picoseconds, after which the potential energy decreases abruptly by several thousand eV over tens of femtoseconds. This is accompanied by a large negative pressure excursion and ultimately by atom loss. Such behavior is consistent with unstable extrapolation of the polynomial MTP representation into poorly sampled atomic environments, where the fitted energy surface can develop unphysical low-energy regions. The present simulations identify the conditions under which this failure mode becomes appreciable, although they do not by themselves establish a unique microscopic origin for the instability.

The transition from stable to unstable behavior is probabilistic rather than a sharply defined density threshold. Transient excursions in temperature were also observed without subsequent failure; at 3.05~g/cm$^3$, for example, 7 of 12 identified excursions recovered without catastrophic instability. The failure fractions reported above therefore apply to the particular system size and 100~ps observation window used here and should not be interpreted as intrinsic failure probabilities independent of simulation duration. For a rare-event instability, longer trajectories or larger systems would generally provide additional opportunities to encounter problematic configurations.

Importantly, the instability is distinct from the structural accuracy measured by the RDFs. MTP reproduces the DFT pair correlations closely at all densities examined in Figure~\ref{fig:rdfs}, including 3.2~g/cm$^3$, while the shorter 216-atom RDF trajectories did not encounter a failure. The potential can therefore provide an accurate description of typical liquid configurations while still exhibiting rare numerical instabilities during substantially longer and larger simulations. The constant-volume tests provide a practical estimate of the region in density--temperature space over which such events become increasingly likely; simulations in other ensembles may remain robust when they sample similar configurations, but density and temperature fluctuations can alter the probability of accessing the unstable regime.

We also calculated the melting temperature of 3C-SiC using TGAP and MTP with the liquid--crystal coexistence method, as described in the Methods section. In these coexistence simulations, we additionally observed separation of the carbon-rich component in the liquid phase, which is discussed below. The melting temperatures predicted by TGAP and MTP are 2600 and 2625~K, respectively, each determined to within 25~K. These values are close to the \textit{ab initio} estimate of $2678.54 \pm 41.67$~K reported in Ref.~\cite{fernan-liq}. They are lower than the available experimental values, namely 2840~K from Ref.~\cite{exp-liq-5} and $2818 \pm 40$~K from Refs.~\cite{fernan-liq,exp-liq-4}, but remain within approximately 200--240~K of experiment.

We next consider the decomposition of carbon in liquid SiC. Here, decomposition refers to separation of carbon from the homogeneous liquid, leading to the formation of carbon-rich clusters or extended carbon phases within a Si-rich liquid matrix~\cite{exp-liq-1,exp-liq-5}. This physical phenomenon should be distinguished from the spurious single-species clustering that can occur in MLIP simulations of disordered phases when important barriers on the potential-energy surface are not adequately represented~\cite{volker-C17,volker-silica,junlei-gao,GeSbTe-volker,guerb2023}. During early stages of TGAP and MTP development, such unphysical carbon clustering was indeed observed in melt--quench simulations. This behavior was mitigated by augmenting the training data with sufficient AIMD snapshots spanning both high- and lower-temperature disordered configurations. In addition, configurations exhibiting clustered carbon environments encountered during iterative TGAP-driven simulations and MTP active-learning cycles were added explicitly to the dataset. The resulting TGAP and MTP models no longer show immediate spurious clustering and reproduce the DFT radial distribution functions shown in Figure~\ref{fig:rdfs}.

Experimental studies have reported decomposition of liquid SiC at atmospheric pressure~\cite{exp-liq-4,exp-liq-5}, making this behavior an important test of whether an interatomic potential can describe the disordered phase beyond homogeneous liquid structure alone. The AIMD coexistence simulations reported in Ref.~\cite{fernan-liq} did not show phase separation, and previously studied classical potentials likewise failed to reproduce carbon decomposition at high temperature and low pressures in the 0--10~GPa range~\cite{Tersoff-SiC-amorph,GW-SiC-amorph}. Among the other MLIPs considered in the literature, UF3, tabGAP, and FLARE-26 have been reported to capture this phenomenon.

The carbon separation observed in our coexistence simulations therefore provides an additional application-level test of the developed potentials.

\begin{figure}[htbp!]
	\centering
	\includegraphics[width=1.0\columnwidth]{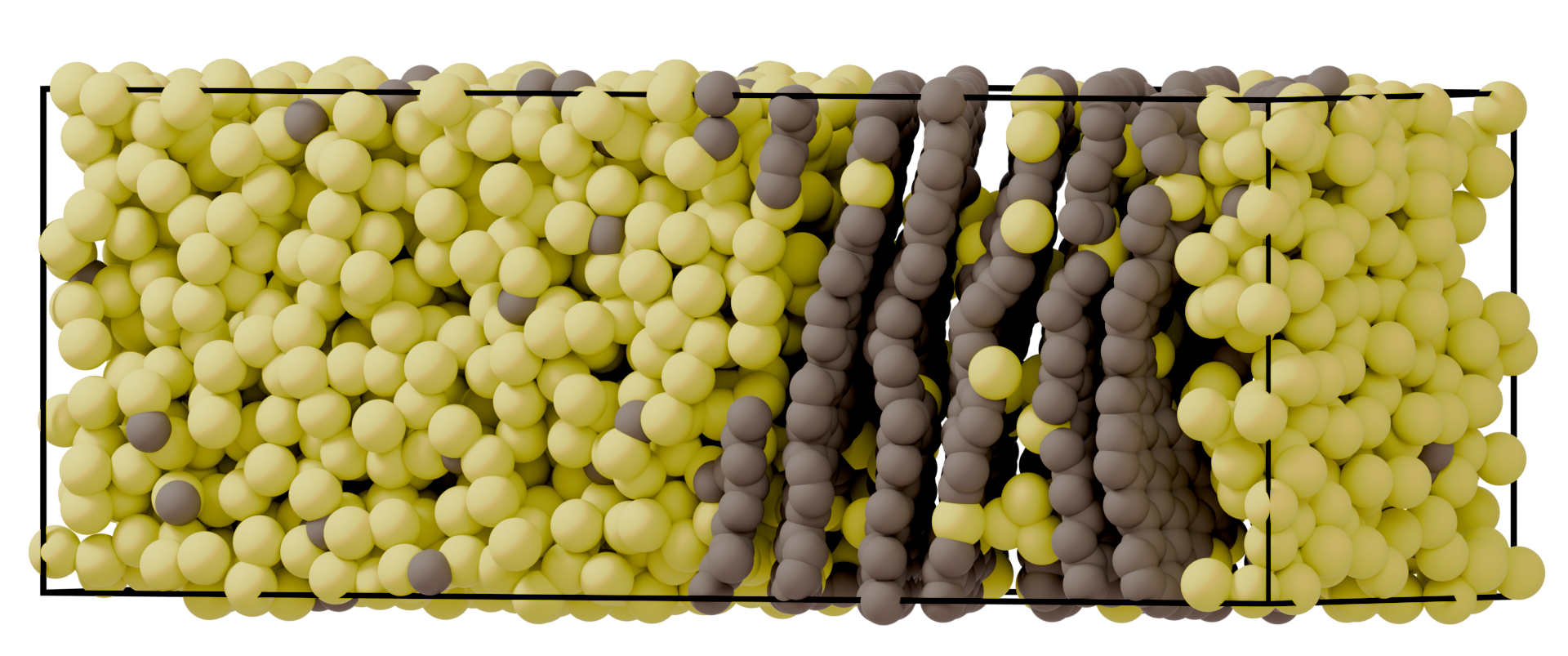}
	\caption{Graphitization of the separated carbon-rich phase observed in the MTP liquid--crystal coexistence simulation at zero pressure. Silicon atoms are shown in yellow-beige and carbon atoms in grey.}
	\label{fig:graphitization}
\end{figure}

In the zero-pressure coexistence simulations, MTP predicts clear graphitization of the separated carbon-rich phase, as illustrated in Figure~\ref{fig:graphitization}. TGAP also predicts carbon separation, but a fully graphitized carbon phase was not observed. Instead, the separated carbon atoms form clusters, the largest of which adopts a distorted graphene-like structure (Supplementary Figure~12). To further assess this behavior, we performed an additional TGAP melt--quench simulation and analyzed the evolution of the carbon clusters during the trajectory. These results are presented in Supplementary Note~11.

\section*{\label{sec:damage}Primary damage in 3C-SiC}

Radiation damage is a multistage process that samples both near-equilibrium defect configurations and strongly non-equilibrium atomic environments. Molecular dynamics simulations of displacement cascades therefore require an interatomic potential capable of describing crystalline and disordered configurations, point defects and defect clusters, and short-range repulsive interactions \cite{hamedani2020insights,hamedani2021primary,byggmastar2019machine}. Previous cascade studies have shown that the earliest stages involve extreme local pressures and short interatomic separations and that the treatment of this regime can directly influence primary defect production~\cite{beland2016primary,beland2016extreme,beland2017shortrange}. Following the initial ballistic collision sequence generated by the primary knock-on atom (PKA), a transient thermal-spike region develops in which the local structure can become strongly disordered or melt-like~\cite{kai-damage}. Subsequent cooling, defect recombination, and recrystallization determine the surviving primary-damage state, which may contain isolated point defects, antisites, small defect clusters, homonuclear bonds, and locally disordered regions.

We simulated displacement cascades initiated by 2~keV PKAs using TGAP and MTP, with the simulation protocol described in the Methods section. These calculations represent isolated primary-damage events rather than cumulative irradiation or amorphization. Nevertheless, they probe the defect-production and recovery processes that precede damage accumulation under repeated irradiation. Previous cascade-overlap simulations have identified defect clustering as an early stage of amorphization, with interstitial--antisite aggregates proposed as possible nuclei for subsequent amorphous growth~\cite{weber2010clustering}. Analysis of defect recombination, chemical disorder, and cluster formation in individual cascades therefore provides an application-level test of the potentials across several configurational regimes relevant to radiation damage.

\begin{figure*}[ht]
	\centering
	\includegraphics[width=0.9\textwidth]{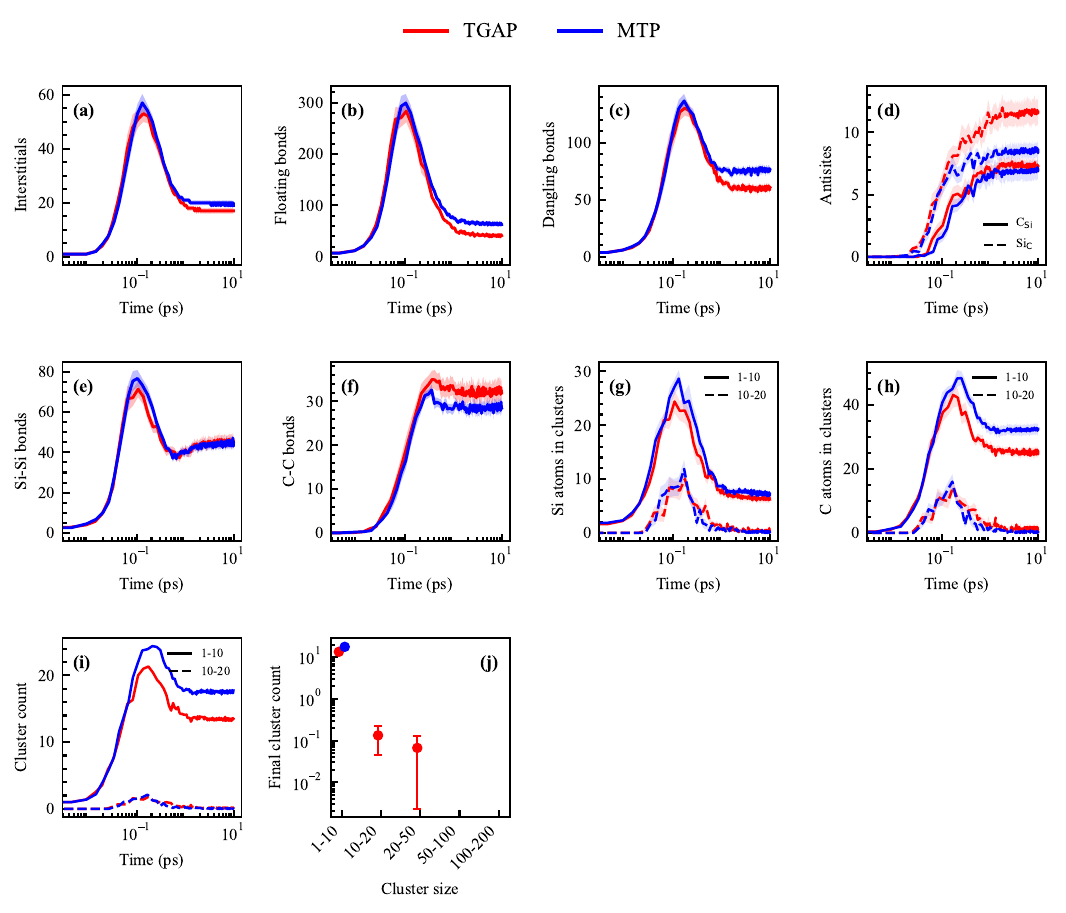}
	\caption{Evolution of the primary-damage state in 3C-SiC following 2~keV PKA cascades calculated with TGAP and MTP and averaged over 20 randomly oriented PKA trajectories. Shaded regions and error bars denote the standard error of the mean over the 20 trajectories. Panel (j) summarizes the surviving defects at the end of the simulations.}
	\label{fig:cascade}
\end{figure*}

Figure~\ref{fig:cascade} summarizes the time evolution of the primary-damage state, averaged over twenty independently oriented 2~keV PKA trajectories.

The interstitial population in Figure~\ref{fig:cascade}(a) shows the characteristic rise and recovery associated with the ballistic and thermal-spike stages. The number of interstitials reaches a maximum near 0.1~ps, with average peak populations of approximately 50 for TGAP and 57 for MTP, before rapidly decreasing to surviving populations of approximately 18 and 19, respectively. The two potentials therefore predict similar overall defect production and recovery at this PKA energy. The rapid approach to the final defect population within the first few picoseconds is also consistent with previous cascade simulations of SiC~\cite{gao2001primary}.

The coordination analysis in Figures~\ref{fig:cascade}(b) and \ref{fig:cascade}(c) provides a complementary measure of the transient structural disorder. Both over-coordinated (``floating'') and under-coordinated (``dangling'') atoms increase sharply during the thermal spike as the local tetrahedral structure is disrupted~\cite{kelires1992short}. Their recovery is asymmetric. With TGAP, the average number of over-coordinated atoms decreases from a peak of approximately 270 to approximately 40, whereas the under-coordinated population decreases from approximately 125 to approximately 60. MTP shows a similar pattern, with corresponding changes from approximately 300 to 65 and from 135 to 75. MTP therefore retains somewhat more coordination disorder after the cascade than TGAP. The preferential persistence of under-coordinated environments is qualitatively consistent with previous studies identifying stable low-coordination Si environments and the tendency of carbon to adopt threefold coordination in disordered SiC~\cite{jin2011driving,kelires1992short}.

The antisite populations in Figure~\ref{fig:cascade}(d) show that Si$_\mathrm{C}$ defects are more numerous than C$_\mathrm{Si}$ defects for both potentials. The final C$_\mathrm{Si}$ populations are similar, averaging 7.7 for TGAP and 7.0 for MTP, whereas the Si$_\mathrm{C}$ populations differ more substantially, averaging 13.0 and 8.7, respectively. Several factors may contribute to this asymmetry. Carbon has the lower orientation-averaged TDE (Table~\ref{tab:ave-tde}), increasing its susceptibility to displacement, while Si$_\mathrm{C}$ also has the lower formation energy of the two antisites (Table~\ref{tab:efs}). The final populations, however, result from the full collision and recombination sequence and cannot be inferred from either quantity alone. The survival of antisites is relevant to damage accumulation because chemical disorder generated during imperfect recrystallization has been identified as an impediment to complete lattice recovery~\cite{malerba2001md}.

Homonuclear Si--Si and C--C bonds are also generated during the cascades, as shown in Figures~\ref{fig:cascade}(e) and \ref{fig:cascade}(f). Both potentials predict more Si--Si than C--C bonds in the surviving damage state. The final Si--Si populations are similar, averaging approximately 49 for TGAP and 45 for MTP, whereas TGAP retains somewhat more C--C bonds, approximately 34 compared with 29 for MTP. Homonuclear bonding provides a measure of chemical disorder in SiC, and previous simulations have associated increasing chemical disorder with the progression toward irradiation-induced amorphization~\cite{yuan2002modeling,leide2019role}. The relative populations of Si--Si and C--C bonds are also known to depend on the interatomic potential~\cite{yuan2002modeling,leide2019role}; the differences observed here therefore provide an additional measure of how TGAP and MTP describe the recovery of the transiently disordered cascade region.

The cluster analysis on interstitials in Figures~\ref{fig:cascade}(g)--\ref{fig:cascade}(j) shows that the surviving damage is dominated by small clusters containing 1--10 defects for both potentials, consistent with previous simulations of individual low-energy cascades in SiC~\cite{gao2001primary,weber2010clustering}. These small clusters are strongly enriched in carbon, containing approximately four to five times as many C as Si atoms in the 1--10-defect size range. MTP also produces more such clusters than TGAP, with averages of approximately 18 and 13 clusters per cascade, respectively. The carbon enrichment is consistent with the lower displacement threshold of carbon and with previous simulations showing that carbon interstitials can form relatively stable clustered or chain-like configurations that inhibit complete recombination~\cite{gao2001primary,jin2011driving,jiang2013amorph}. Similar defect aggregates have been implicated in the early stages of damage accumulation during cascade overlap~\cite{weber2010clustering}, although the present single-cascade simulations do not establish their subsequent evolution into amorphous regions.

Taken together, the 2~keV cascade simulations provide a demanding application-level test of TGAP and MTP under conditions that sequentially sample short-range repulsion, highly disordered local environments, recrystallization, and residual defect structures. The two potentials give similar predictions for the overall magnitude and timescale of primary defect production and recovery, while differing in the amount of residual coordination disorder, antisite production, homonuclear bonding, and small-cluster formation. These calculations should not be interpreted as a validation of cumulative irradiation damage or amorphization, for which cascade-overlap simulations and comparison with independent experimental or higher-level reference data would be required. Rather, they demonstrate that both potentials remain stable and produce physically plausible primary-damage evolution in isolated cascades, providing a basis for future simulations of damage accumulation at larger doses.

\section*{\label{sec:efficiency}Computational efficiency and large-scale simulation}

The accuracy of an interatomic potential determines the range of properties and configurations it can describe, whereas its computational cost determines the system sizes and time scales that are practically accessible. The balance between these quantities is therefore an important consideration when selecting a model for a particular application. We benchmarked the computational performance of the MLIPs for which compatible LAMMPS implementations were used in this work under a common simulation protocol. A pristine 3C-SiC supercell was propagated for 2000 steps in the microcanonical ensemble at 300~K with a 1~fs timestep, using the same initial velocity seed for each potential. Each benchmark was repeated three times, and the lowest measured cost is reported as an estimate of the uncontended execution time. Because per-atom computational cost depends on system size, hardware, implementation, and parallel efficiency, the values reported here are intended primarily for comparison among models under the specified benchmark conditions.

The CPU benchmark used a 1000-atom supercell running on a single core of an AMD EPYC 7713 processor at 2.0~GHz. The measured costs are listed in Table~\ref{tab:efficiency}.

\begin{table}[!htbp]
	\centering
	\begin{threeparttable}
		\caption{Computational cost of the MLIPs in the single-core CPU benchmark. A 1000-atom 3C-SiC supercell was propagated for 2000 NVE steps at 300~K on one AMD EPYC 7713 core. Each value is the lowest of three repeats; the spread across repeats ranged from 0.5 to 38\%. Relative costs are normalized to the fastest model under these conditions.}
		\label{tab:efficiency}
		\renewcommand{\arraystretch}{1.3}
		\begin{tabular}{lcc}
			\toprule
			Potential & Cost ($\mu$s$\cdot$core/atom/step) & Relative cost \\
			\midrule
			tabGAP   & 49    & 1.0 \\
			FLARE-26 & 113   & 2.3 \\
			MTP      & 200   & 4.1 \\
			UF3      & 253   & 5.1 \\
			DP-GB    & 652   & 13.2 \\
			DP-ZBL   & 2174  & 44.1 \\
			TGAP     & 2549  & 51.8 \\
			DP-TH    & 4656  & 94.5 \\
			\bottomrule
		\end{tabular}
	\end{threeparttable}
\end{table}

The measured CPU costs span almost two orders of magnitude, from 49 to 4656~$\mu$s$\cdot$core/atom/step. LAMMPS timing data attribute more than 99.9\% of the runtime in these benchmarks to the force-evaluation portion of the calculation, indicating that differences among the models dominate the measured performance. The substantial variation among models using the same general MLIP framework also shows that computational cost is determined not only by the underlying formalism, but by the specific model architecture and implementation. For example, the three Deep Potential models differ by approximately a factor of seven in CPU cost.

TGAP and MTP occupy substantially different positions on this accuracy--cost spectrum. Under the single-core benchmark conditions, MTP evaluates forces approximately 12.8 times faster than TGAP. Its cost is comparable to that of UF3 and approximately four times that of tabGAP, the fastest model tested. TGAP is substantially more expensive, reflecting the higher computational cost of its descriptor evaluation and regression scheme. The two developed potentials therefore provide distinct options depending on whether greater emphasis is placed on configurational accuracy or computational throughput.

The MTP implementation used here also supports GPU acceleration through our Kokkos implementation~\cite{mtp-kokkos}. The custom MTP basis used in this work was optimized primarily for CPU inference, however, and therefore does not attain the same GPU acceleration as some more conventional MTP constructions. We repeated the performance benchmark on a single NVIDIA H100 SXM5 GPU using a 32768-atom supercell. This system size was chosen because it could be executed with all of the GPU-capable models included in the comparison; at 110592 atoms, DP-TH and DP-ZBL exceeded the available device memory. TGAP is not included because the \verb|soap_turbo| implementation used here does not support GPU execution through LAMMPS. Run-to-run variation was below 3\% for all models except UF3.

The GPU results are summarized in Table~\ref{tab:efficiency-gpu}.

\begin{table}[!htbp]
	\centering
	\begin{threeparttable}
		\caption{Computational cost of the GPU-capable MLIPs on a single NVIDIA H100 SXM5 GPU using a 32768-atom 3C-SiC supercell. Each value is the lowest of three repeats; the spread across repeats ranged from 0 to 7\%. Relative costs are normalized to the fastest model under these conditions. TGAP is omitted because the \texttt{soap\_turbo} implementation used here does not support GPU execution in LAMMPS.}
		\label{tab:efficiency-gpu}
		\renewcommand{\arraystretch}{1.3}
		\begin{tabular}{lcc}
			\toprule
			Potential & Cost ($\mu$s/atom/step) & Relative cost \\
			\midrule
			tabGAP   & 0.096 & 1.0 \\
			FLARE-26 & 0.132 & 1.4 \\
			MTP      & 0.899 & 9.4 \\
			DP-GB    & 1.152 & 12.0 \\
			UF3      & 1.155 & 12.0 \\
			DP-TH    & 7.508 & 78.1 \\
			DP-ZBL   & 8.494 & 88.4 \\
			\bottomrule
		\end{tabular}
	\end{threeparttable}
\end{table}

The GPU ordering broadly resembles the CPU benchmark, with tabGAP and FLARE-26 remaining the fastest models and MTP ranking third. DP-GB and UF3 have essentially identical costs under the present GPU conditions, while the relative ordering of DP-TH and DP-ZBL is reversed compared with the CPU benchmark. For MTP, the ratio between the single-core CPU and single-GPU per-atom costs is approximately 220. This ratio provides an indication of the acceleration obtained with the GPU implementation under the respective benchmark configurations, but should not be interpreted as a hardware-independent equivalence between one H100 GPU and 220 CPU cores because the CPU and GPU tests use different system sizes and parallelization regimes. tabGAP and FLARE-26 had not yet reached their asymptotic GPU throughput at 32768 atoms; their measured throughputs therefore represent lower bounds on the peak throughput attainable at larger system sizes.

To demonstrate the system sizes accessible with MTP in a production calculation, we simulated a 40~keV displacement cascade in a $120\times120\times120$ 3C-SiC supercell containing 13,824,000 atoms, following the same general cascade protocol used for the 2~keV simulations. The calculation was performed on 1920 CPU cores and completed in 6~h~11~min, corresponding to an average throughput of 13.7~million atom-steps per second. This is approximately 7100 atom-steps/s/core, compared with approximately 5000 atom-steps/s in the 1000-atom single-core benchmark. Each core carries a substantially larger local workload in the cascade calculation, approximately 7200 atoms, allowing fixed computational overheads to be amortized more effectively. These measurements demonstrate efficient execution at this scale, although a dedicated strong- or weak-scaling analysis would be required to quantify parallel scaling efficiency.

\begin{figure}[htbp!]
	\centering
	\includegraphics[width=1.0\columnwidth]{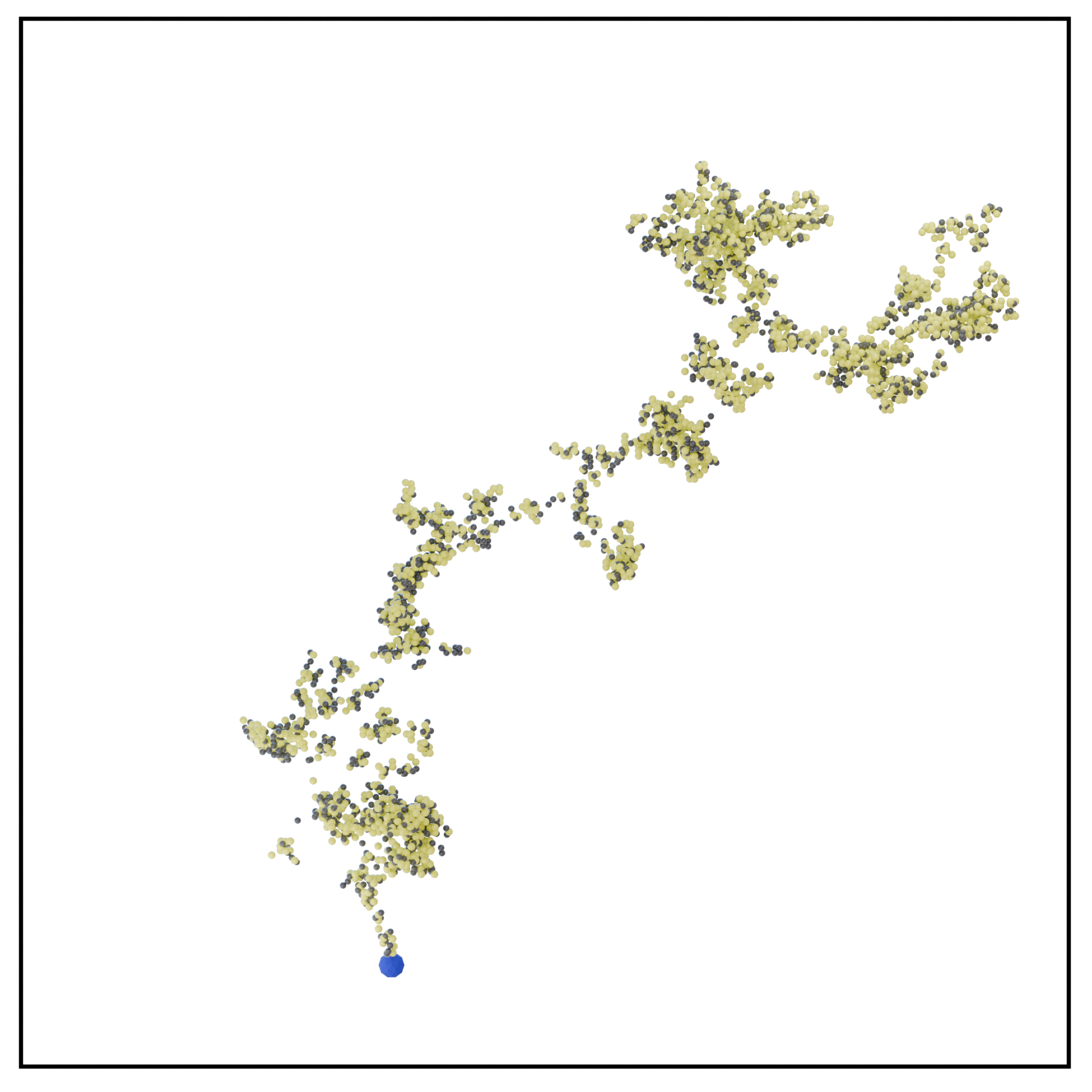}
	\caption{Snapshot of a 40~keV displacement cascade in 3C-SiC near the thermal-spike stage, simulated with MTP in a $120\times120\times120$ supercell containing 13,824,000 atoms. Atoms with kinetic energies above 0.48~eV are shown. The simulation was performed on 1920 CPU cores and completed in 6~h~11~min. The blue marker indicates the initial PKA position. Silicon atoms are shown in yellow-beige and carbon atoms in grey. The simulation cell is viewed from above.}
	\label{fig:large_cascade}
\end{figure}

As shown in Figure \ref{fig:large_cascade}, the 40~keV cascade develops a dense, highly excited region that subsequently separates into two principal branches, with smaller branches extending along the collision path. The purpose of this calculation is primarily to demonstrate the computational scale accessible with MTP rather than to provide a statistical characterization of 40~keV primary damage, for which multiple independent cascade trajectories would be required.

Taken together, the performance benchmarks clarify the complementary roles of the two potentials developed here. TGAP provides the more computationally intensive description and, with the present implementation, is restricted to CPU execution. MTP is approximately an order of magnitude less expensive on a single CPU core, additionally supports GPU acceleration, and enables simulations containing tens of millions of atoms. These differences illustrate the central accuracy--cost tradeoff of MLIP selection: the appropriate model depends not only on the properties that must be reproduced, but also on the spatial and temporal scales required for the intended application.

\section*{Discussion}

We have developed two complementary machine-learning interatomic potentials for 3C-SiC using DFT-labeled datasets spanning crystalline, liquid, and amorphous configurations together with 21 types of point defects and defect complexes. TGAP, based on the GAP framework with the \verb|soap_turbo| descriptor, was designed to provide high accuracy across a broad range of atomic environments, whereas MTP was designed to retain good accuracy at substantially lower computational cost and with access to GPU acceleration. Both potentials are coupled to the NLH short-range repulsive interaction, allowing the same models to describe near-equilibrium bonding and the strongly repulsive configurations encountered during energetic collisions. In parallel, we benchmarked seven existing MLIPs and selected empirical potentials using a common set of structural, mechanical, thermal, defect, disordered-phase, radiation-damage, and computational-performance tests. This common benchmark makes clear that no single model is uniformly optimal across all properties and instead exposes the accuracy--cost and configurational-coverage tradeoffs relevant to model selection.

For near-equilibrium properties, TGAP and MTP both reproduce the DFT description of 3C-SiC well, although their relative performance varies by observable. MTP gives particularly close agreement with the DFT elastic constants, while both potentials reproduce the equilibrium lattice constant, energy--volume relation, and phonon dispersion accurately. Their finite-temperature behavior is also generally good, but the comparison highlights an important distinction between agreement with DFT and agreement with experiment. Both models reproduce the overall thermal softening of Young's modulus, whereas the individual elastic constants show component-dependent differences: in particular, none of the tested MLIPs reproduces the comparatively weak experimental temperature dependence of $C_{44}$. Within the quasi-harmonic approximation, TGAP and MTP closely reproduce the DFT heat capacity and entropy, while MTP gives the closest agreement with the available experimental thermal-expansion data. These results illustrate that agreement with a single reference or equilibrium property is insufficient to establish uniform accuracy across the finite-temperature elastic and vibrational response.

The defect benchmarks provide a more discriminating comparison. TGAP reproduces the DFT defect formation energies and their relative ordering most consistently across the configurations considered, while MTP captures many of the same energetic trends and relaxation pathways but tends to underestimate the formation energies of higher-energy defects. Both models reproduce several unstable interstitial transformations observed in DFT. Vacancy migration further illustrates the property-dependent nature of the benchmark: TGAP most accurately reproduces the C-vacancy second-nearest-neighbor barrier, whereas MTP and tabGAP perform better for the corresponding Si-vacancy hop. The $V_\mathrm{Si}\text{-}\mathrm{Si}_\mathrm{C}\rightarrow V_\mathrm{Si}$ vacancy--antisite pathway, which is not represented explicitly in the training dataset, is more challenging for all of the models and provides a useful test beyond the fitted defect configurations. The larger discrepancies for this pathway emphasize the distinction between accuracy within the sampled configurational domain and transfer to previously unrepresented local environments.

Threshold displacement energies provide a complementary test of the models under strongly non-equilibrium conditions. TGAP and MTP reproduce several of the available low-index experimental and \textit{ab initio} values, although the reference data themselves show substantial variation with crystallographic direction, computational methodology, and experimental technique. Sampling 154 recoil directions further reveals significant anisotropy and greater model dependence for Si than for C recoils. TGAP and MTP predict orientation-averaged TDEs of 37--41~eV for Si and 24--26~eV for C, while the minimum sampled thresholds are considerably lower. Together with tabGAP, the results suggest a characteristic minimum directional threshold of approximately $17\pm1$~eV for Si, whereas all four MLIPs considered in the angular analysis give a minimum of approximately 14~eV for C. These minimum values should be distinguished from the orientation-averaged TDEs and may be useful when parameterizing reduced radiation-damage models in which a minimum displacement threshold is required; the appropriate choice depends on how directional anisotropy is represented.

The disordered-phase calculations similarly reveal both strengths and limitations that are not apparent from equilibrium benchmarks alone. TGAP and MTP reproduce the DFT liquid radial distribution functions across the sampled density range and predict melting temperatures of 2600 and 2625~K, respectively. These values lie close to the reported \textit{ab initio} melting temperature of approximately 2679~K but are roughly 200~K below the available experimental estimates near 2820--2840~K. Both potentials also capture carbon separation from liquid SiC. MTP proceeds further to form a graphitized carbon-rich phase, consistent with the experimentally observed decomposition behavior, whereas TGAP produces carbon-rich clusters with distorted graphene-like local structure but does not fully graphitize on the simulated time scale.

Accurate short trajectories in the liquid do not necessarily guarantee long-time numerical robustness. Extended high-temperature simulations with MTP reveal a rare failure mode, not observed with TGAP, whose probability increases as the liquid is compressed toward the upper end of the sampled density range. The onset of these events is consistent with the polynomial model entering poorly sampled local environments and developing an unphysical low-energy trajectory, although the present calculations do not uniquely establish the microscopic origin. This limitation is important for large-scale applications: MTP can reproduce typical liquid structure accurately while still encountering rare unstable configurations over sufficiently long trajectories or large system sizes. Such behavior highlights the importance of evaluating not only conventional test errors and structural observables, but also stability over the exposure times and configurational regimes relevant to the intended simulation.

The 2~keV displacement cascades provide an application-level test in which several of these regimes are encountered sequentially within a single trajectory, including short-range repulsion, strong local disorder, recrystallization, and residual defect formation. TGAP and MTP predict similar peak and surviving defect populations and comparable overall recovery times, while differing in the amount of residual coordination disorder, antisite production, homonuclear bonding, and small-cluster formation. In the absence of an independent first-principles or experimental reference for these specific cascade observables, these calculations should not be regarded as quantitative validation of cascade damage or amorphization. Rather, they show that both potentials remain stable during isolated 2~keV cascades and produce physically plausible primary-damage evolution consistent with known features of SiC irradiation. Extending these calculations to cascade overlap and accumulated dose will be necessary to determine how differences between the models propagate into predictions of amorphization and long-time damage evolution.

The computational benchmarks further emphasize that accuracy and accessible simulation scale must be considered together. On a single CPU core, MTP is approximately thirteen times faster than TGAP and approximately four times more expensive than tabGAP, the fastest model in the benchmark. MTP additionally supports GPU execution through the implementation used here, whereas the present LAMMPS implementation of TGAP does not. A 40~keV MTP cascade containing 13.8~million atoms was completed in approximately six hours on 1920 CPU cores, demonstrating that the model can access the system sizes required for high-energy cascade simulations. This combination of broad configurational coverage, comparatively high accuracy, and computational efficiency distinguishes MTP from the other models considered here.

More broadly, the benchmark indicates that TGAP and MTP represent a meaningful advance in interatomic modeling of 3C-SiC. In the case of MTP, the present model extends a development trajectory in which increasingly broad configurational and chemical domains have been combined within computationally efficient potentials~\cite{luo2023mtp,zongo2024unified,sun2024interatomic,zongo2025amorphous,zongo2026machine}. TGAP provides consistently strong agreement with DFT across equilibrium, vibrational, defect, and disordered configurations and reproduces most of the defect relaxation pathways examined here. MTP retains much of this accuracy while reducing the single-core computational cost by more than an order of magnitude, providing GPU support, and reproducing additional demanding behavior such as carbon separation and graphitization in the liquid phase. Although individual existing potentials outperform TGAP or MTP for particular observables, none provides the same overall combination of configurational coverage, accuracy across the present benchmark, and computational accessibility.

On this basis, we conclude that TGAP and MTP have their merits within their natural domains of application. MTP is particularly well suited to large-scale simulations of 3C-SiC in which mechanical response, defect evolution, melting, decomposition, or radiation damage must be treated within a single potential. It combines broad configurational coverage and close agreement with DFT and experiment for several key properties with the computational efficiency and GPU support required for multi-million-atom simulations, and it reproduces the graphitization of carbon-rich regions observed experimentally during decomposition. For simulations involving prolonged sampling of high-density liquid configurations, the MTP extrapolation criterion should be monitored and the potential used within an active-learning workflow, with extrapolative configurations incorporated into the training set as necessary. Such strategies are well established within the MTP framework, and our previous work has shown that targeted active learning and small-cell training can substantially reduce the reference-data cost required to extend an MLIP into new configurational regimes~\cite{meziere2023accelerating,meng2025smallcell}. TGAP is the natural choice when accuracy is the priority and the required system sizes are moderate, particularly for defect studies, where it reproduces the DFT formation energies and their relative ordering most consistently among the models considered, including those of higher-energy defects.

Taken together, the two potentials extend the capabilities available for atomistic simulation of 3C-SiC. TGAP provides a high-accuracy model with broad configurational coverage, while MTP offers a particularly favorable compromise between accuracy, computational cost, and accessible system size. The systematic comparison with existing potentials further shows that model selection remains application dependent, but the breadth of the present benchmark supports MTP as our preferred model for many large-scale and strongly non-equilibrium applications. The combination of these models and the benchmark presented here therefore provides a practical foundation for simulations ranging from equilibrium materials properties to melting, decomposition, defect evolution, and radiation damage.

\section*{Methods}

Supplementary Notes~1 and 2 describe the construction of the training datasets and the selection of fitting hyperparameters, respectively. Numerical validation of the fitted potentials is provided in Supplementary Note~3. Computational details of the DFT labeling calculations and additional validation simulations are given in Supplementary Note~4, and the formulation and parametrization of the NLH short-range repulsive potential are described in Supplementary Note~7.

\textbf{TGAP training.}
The GAP framework and the \verb|soap_turbo| descriptor are described in Refs.~\cite{volker-review} and \cite{caro2019optimizing}, respectively. Reference energies and atomic forces were calculated with VASP using the projector augmented-wave (PAW) method~\cite{paw1} and the Perdew--Burke--Ernzerhof (PBE) generalized-gradient approximation for exchange and correlation. Additional DFT settings are provided in Supplementary Note~4. TGAP was fitted using the \verb|gap_fit| code distributed with QUIP~\cite{QUIP}. MLIP-driven molecular dynamics simulations were performed with TurboGAP and LAMMPS, with GAP calculations in LAMMPS interfaced through QUIP.

\textbf{MTP training.}
MTP training was performed using a modified version of MLIP-3~\cite{podryabinkin2023mlip}. The implementation was extended to support the custom radial envelope function used in the present model, together with modifications to the numerical routines required to efficiently train the enlarged basis set. MLIP-driven molecular dynamics simulations were performed using custom LAMMPS implementations supporting the modified radial basis and short-range hybridization. A custom implementation of the NLH repulsive interaction was used in the present calculations, although the same tabulated interaction can alternatively be supplied through the native LAMMPS tabulated-pair interface. The relevant implementations are available from the repositories cited in Refs.~\cite{lammps_mtp_github,mtp_kokkos_github}.

\textbf{Threshold displacement energy calculations.}
For each TDE calculation, the pristine supercell was first equilibrated for 15~ps at the target temperature and zero pressure in the isothermal--isobaric (NPT) ensemble. During the recoil simulation, atoms within a 2~\AA-thick layer at each boundary of the simulation cell were thermostatted using a Nos\'e--Hoover thermostat, while the remaining atoms evolved using no thermostat. The simulation cell was sufficiently large that recoils with energies up to 160~eV did not interact with the thermostatted boundary region.

An adaptive timestep was used during the recoil simulations. The maximum atomic displacement permitted within a single integration step was 0.015~\AA, and the timestep was constrained not to exceed 3~fs. Both Si and C atoms were considered as PKAs. For each recoil direction, the search began at a PKA energy of 10~eV. After 1.5~ps, the configuration was analyzed for defects using Wigner--Seitz analysis as implemented in OVITO~\cite{ovito}. If no defect was detected, the calculation was repeated with the PKA energy increased by 2~eV. If a defect was present at 1.5~ps, the trajectory was continued to 6~ps and analyzed again to determine whether the defect survived. This procedure was repeated until the lowest recoil energy producing a defect that remained present at 6~ps was identified. For the low-index TGAP calculations, the resulting threshold was further refined using increments of 0.5~eV. All TDE simulations were performed with LAMMPS.

\textbf{Melting-temperature calculations.}
The melting temperature was determined using a solid--liquid coexistence method. A non-cubic $6\times6\times12$ supercell containing 3456 atoms was used. For each trial temperature, $T$, the initial crystalline cell was equilibrated for 10~ps in the NPT ensemble at zero pressure, with the three cell dimensions allowed to respond independently. The atoms in the upper half of the cell along the $z$ direction were then immobilized, and the lower half was melted at 6000~K for 20~ps in the canonical (NVT) ensemble. The molten half was subsequently equilibrated for an additional 10~ps at the trial temperature in the NPT ensemble, with pressure control applied along the $z$ direction.

The constraint on the crystalline half was then removed, and the combined solid--liquid system was evolved at the trial temperature and zero pressure for at least 1~ns. The temperature at which neither phase consumed the other over this interval was identified as the coexistence temperature and hence the melting temperature. The initial temperature search began at 3250~K and proceeded downward in increments of 250~K to bracket the transition. The melting temperature was then refined using 25~K increments. A timestep of 1~fs was used throughout. All coexistence simulations were performed with LAMMPS.

\textbf{Primary-damage simulations.}
For the 2~keV cascade simulations, pristine $20\times20\times20$ supercells containing 64,000 atoms were constructed using the equilibrium lattice constant of the corresponding potential. Each cell was equilibrated for 20~ps at 300~K and zero pressure in the NPT ensemble using a 1~fs timestep. A PKA was then selected from a region near the center of the cell and assigned a 2~keV kinetic energy in a randomly selected direction directed toward the cell interior. Twenty independent trajectories were simulated to sample the variability associated with PKA position and direction.

During the cascade, atoms in a boundary layer of the supercell were thermostatted in the NVT ensemble to remove excess heat from the cascade region, while the interior evolved without thermostatting. Electronic stopping was represented by a non-local friction force applied to atoms above a kinetic-energy threshold of 0.1~eV. An adaptive timestep was used during the cascade. Each trajectory was continued for 10~ps, by which time the simulation cell had returned to approximately 300~K.

Post-processing was performed with OVITO. Defect clusters were identified using a clustering cutoff of 4.38~\AA, approximately equal to the 3C-SiC lattice constant. Atomic coordination was determined using a cutoff of 2.5~\AA, which lies approximately midway between the first- and second-nearest-neighbor distances in 3C-SiC. Homonuclear bonds were identified using cutoff distances of 2.42~\AA\ for Si--Si pairs and 1.55~\AA\ for C--C pairs.

For the 40~keV demonstration calculation, the same general cascade protocol was applied to a pristine $120\times120\times120$ supercell containing 13,824,000 atoms. A single trajectory was simulated on 1920 CPU cores. Computational performance for this calculation is reported in Section~\ref{sec:efficiency}.

\section*{Data availability}

The trained TGAP and MTP potential files and the datasets generated in this study will be deposited on Zenodo and made publicly available upon publication. During peer review, these materials are available to editors and referees upon request. The software implementations used for MTP training and molecular dynamics simulations, including the custom LAMMPS and Kokkos implementations, are publicly available through the GitHub repositories described in Refs.~\cite{lammps_mtp_github,mtp_kokkos_github}.

\section*{Acknowledgments}
The authors acknowledge the computational resources provided by the Aalto Science-IT project and CSC--IT Center for Science, Finland. The authors also acknowledge the Digital Research Alliance of Canada for the generous allocation of computational resources. The views and opinions expressed herein are those of the authors only and do not necessarily reflect those of the European Union, the European Research Council Executive Agency, the Natural Sciences and Engineering Research Council of Canada (NSERC), Mitacs, the University Network for Excellence in Nuclear Engineering (UNENE), or any other funding or supporting organization. Neither the European Union nor the granting authority can be held responsible for them.

\section*{Funding}
This work was supported by the European Union through the European Research Council (ERC-2022-STG, project MUST, No.~101077454). This work was also supported by the Natural Sciences and Engineering Research Council of Canada (NSERC), Mitacs, and the University Network for Excellence in Nuclear Engineering (UNENE).

\subsection*{Author contributions}
A.H. and A.S. conceived the project. A.H. developed the TGAP potential, generated the training data for both potentials, and performed the validation simulations. Z.M. implemented and trained the MTP and its GPU-accelerated version. All authors contributed to the interpretation of the results. A.H. prepared the manuscript, and all authors reviewed and edited it. A.S. and L.K.B. supervised the work. Project administration and funding acquisition were performed by A.S and L.K.B.

\subsection*{Competing Interest}
The authors declare no competing interests.

%

\clearpage

\pdfximage{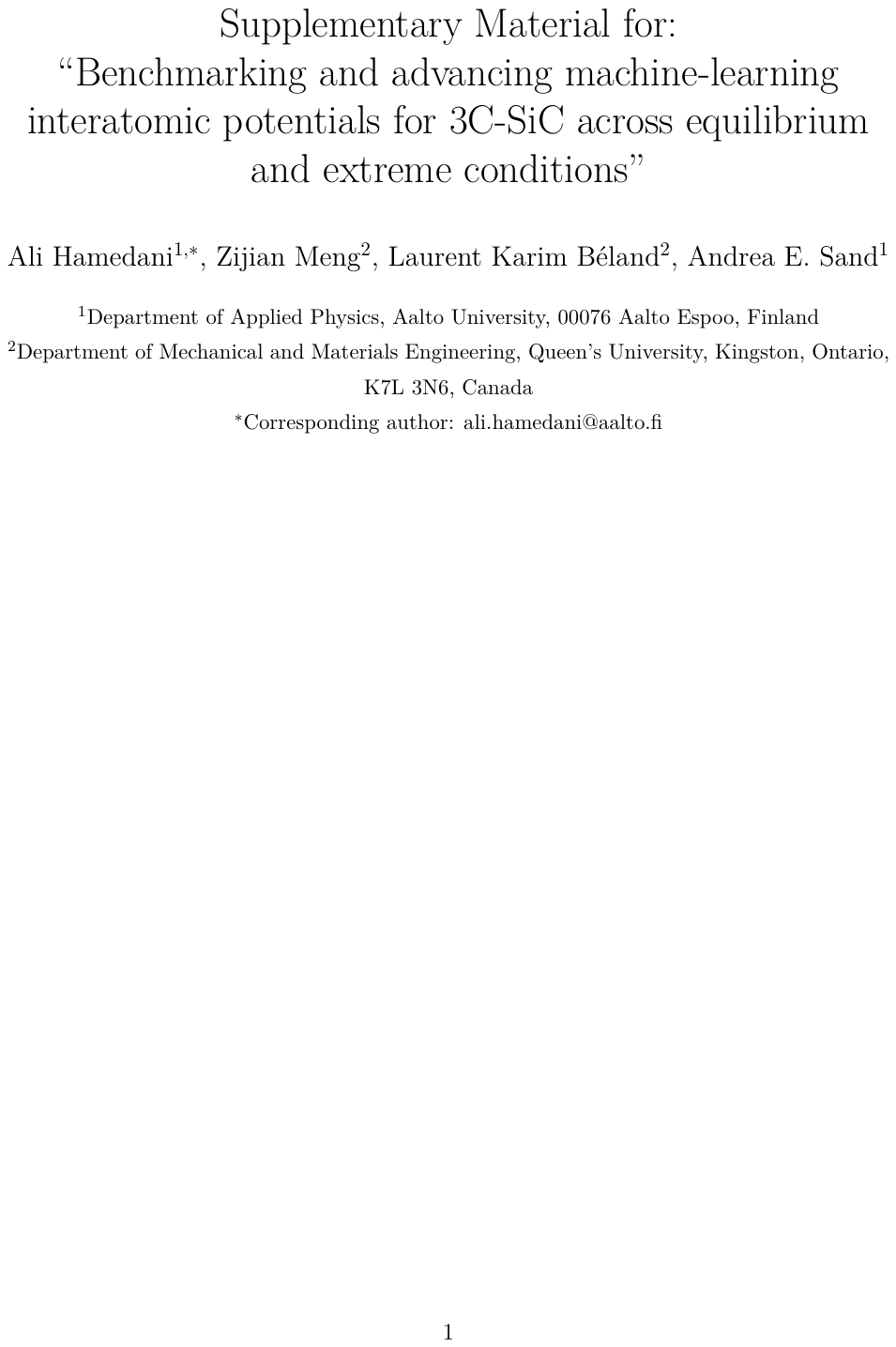}
\edef\supppages{\the\pdflastximagepages}
\pdfpagewidth=210mm \pdfpageheight=297mm   
\pdfhorigin=0pt \pdfvorigin=0pt \hoffset=0pt \voffset=0pt
\newcount\suppage \suppage=1
\loop
\pdfximage page \suppage {supplementary.pdf}%
\shipout\hbox{\pdfrefximage\pdflastximage}%
\advance\suppage by 1
\ifnum\suppage<\numexpr\supppages+1\relax\repeat

\end{document}